\documentclass[twoside,journal]{IEEEtran}

\usepackage[latin1]{inputenc} 
\usepackage{ifthen} 
\usepackage{color} 

\usepackage[font=footnotesize,center]{caption} 
\usepackage[font=footnotesize]{subcaption} 
\usepackage{bm} 
\usepackage[sans]{dsfont} 

\usepackage{graphicx}
\usepackage[cmex10]{amsmath}
\usepackage{amsthm}
\usepackage[noadjust]{cite}
\usepackage{algpseudocode}
\usepackage{url}
\usepackage{wrapfig} 
\graphicspath{{images/}{photos/}{graphics/}{patternsVsNumRebounds/}}

\graphicspath{{images/}{graphics/}{photos/}}

\usepackage{algorithm}
\usepackage{multirow} 
\usepackage{amssymb} 

\newcommand{\mbf}{\mathbf}
\newcommand{\mrm}{\mathrm}

\definecolor{greenao}{rgb}{0.0, 0.5, 0.0}

\newtheorem{definition}{Definition}
\newtheorem{theorem}{Theorem}
\newtheorem{lemma}{Lemma}
\newtheorem{coro}{Corollary}[theorem]
\DeclareMathOperator{\rank}{rank} 
\DeclareMathOperator{\tr}{tr} 
\newcommand{\eg}{e.g.~}

\newcommand{\lparen}{\left(}
\newcommand{\rparen}{\right)}
\newcommand{\mm}{\,}  
\newcommand{\hp}{\,\circ\,}  

\newcommand{\numMics}{n}
\newcommand{\TDOASETn}{\mathcal{M}_{T}(n)}
\newcommand{\TDOAMatrix}{\mbf{M}}

\newcommand{\useAlgNames}{true}

\newcommand{\RobustDenoisingAlg}{\texttt{Robust DeN}}
\newcommand{\MatrixCompletionAlg}{\texttt{MC}}
\newcommand{\RobustDenoisingMatrixCompletionAlg}{\RobustDenoisingAlg+\MatrixCompletionAlg}

\newcommand*{\mathcolor}{}
\def\mathcolor#1#{\mathcoloraux{#1}}
\newcommand*{\mathcoloraux}[3]{%
  \protect\leavevmode
  \begingroup
    \color#1{#2}#3%
  \endgroup
}

\begin{document}

\title{TDOA Matrices: Algebraic Properties and their Application to Robust Denoising with Missing Data}
\author{Jose~Velasco,\thanks{J.~Velasco, D.~Pizarro, and J.~Macias-Guarasa are with the Department of Electronics, Escuela Politécnica
    Superior, University of Alcalá, 28805 Alcalá de Henares, Spain.}
  Daniel~Pizarro,
Javier~Macias-Guarasa
and~Afsaneh~Asaei,\thanks{A.~Asaei is
  with Idiap Research Institute, Martigny, Switzerland.}
\thanks{E-mails: \{jose.velasco, pizarro, macias\}@depeca.uah.es,
  \nobreak afsaneh.asaei@idiap.ch).}
\thanks{Manuscript received Month Day, 2015; revised Month Day, 2015.}}

\maketitle

\begin{abstract}
  Measuring the Time delay of Arrival (TDOA) between a set of sensors
  is the basic setup for many applications, such as localization or
  signal beamforming. This paper presents the set of TDOA matrices,
  which are built from noise-free TDOA measurements, not requiring
  knowledge of the sensor array geometry. We prove that TDOA matrices
  are rank-two and have a special SVD decomposition that leads to a
  compact linear parametric representation. Properties of TDOA
  matrices are applied in this paper to perform denoising, by finding
  the TDOA matrix closest to the matrix composed with noisy
  measurements.  The paper shows that this problem admits a
  closed-form solution for TDOA measurements contaminated with
  Gaussian noise which extends to the case of having missing data. The
  paper also proposes a novel robust denoising method resistant to
  outliers, missing data and inspired in recent advances in robust
  low-rank estimation. Experiments in synthetic and real datasets show
  significant improvements of the proposed denoising algorithms in
  TDOA-based localization, both in terms of TDOA accuracy estimation
  and localization error.
\end{abstract}

\section{Introduction}
%
%
%

\IEEEPARstart{T}{ime} delay of arrival (TDOA) estimation is an
essential pre-processing step for multiple applications in the context
of sensor array processing, such as multi-channel source
localization~\cite{Sheng2005MaximumLikelihood},
self-calibration~\cite{kuang2013stratified} and
beamforming~\cite{anguera2007acoustic}. In all cases, performance is
directly related to the accuracy of the estimated TDOAs
\cite{brandstein01}.

Estimating TDOA in noisy environments has been subject of study during
the last two decades \cite{carter1981special, chen2006time,
  ho2012bias}, and is still an active area of research, benefiting
from current advances in signal processing and optimization
strategies~\cite{alameda2014geometric, compagnoni2014comprehensive,
  nouvellet2014slowness, huang2015tdoa}. 

Typically, the TDOA between a single pair of sensors is obtained by
measuring the peak of the generalized cross-correlation (GCC) of the
received signals on each sensor~\cite{dibiase2001robust}, which are
assumed to be generated from a single source. Many factors, such as the
spectral content of the signal, multipath propagation, and noise
contribute to errors in the estimation of the TDOA. 

Given a set of sensors, TDOA measurements can be obtained for every
possible pair of sensors. This is commonly known as the \emph{full TDOA
  set} or \emph{spherical set}~\cite{yang2006theoretical}. This paper
studies how to reduce noise and errors from the full TDOA set. The
intuition behind this denoising is to exploit redundancy of the full
TDOA set. For $n$ sensors, the full set of $n(n-1)/2$ measurements can
be represented by $n-1$ values, which are referred to as the
\emph{non-redundant set}.  This problem has been studied in the past,
showing that one can optimally obtain the non-redundant set when TDOA
measurements are contaminated with additive Gaussian noise. This is
known as the Gauss-Markov estimator~\cite{hahn73optimun}. However, in
more realistic scenarios errors are not Gaussian and some of the TDOA
measurements may contain outliers. In these cases the Gauss-Markov
estimator performs poorly.

This paper presents the TDOA matrix, which is created by the
arrangement of the full TDOA set inside a skew-symmetric matrix, and
studies the algebraic properties of this matrix, showing that it has
rank $2$ and a SVD decomposition with $n-1$ degrees of freedom. Such
matrices have been previously defined in the
literature~\cite{Zhu2011}, but their properties and applications have
not been studied until now.

These algebraic properties are used in this paper to perform denoising
under different scenarios, which include the presence of missing TDOA
measurements and outliers. These denoising algorithms are tested in
the context of speaker localization with microphone arrays, using
synthetic and publicly available real datasets.  Our denoising
algorithms are able to recover accurate TDOA values for high rates of
missing data and outliers, significantly outperforming the
Gauss-Markov estimator in those cases. All the proposed methods don't
require knowledge of the sensor positions, so that they can also be
used for calibration~\cite{kuang2013stratified}.


The main contributions of this work are threefold: \emph{i)} Definition
of the algebraic properties of TDOA matrices. \emph{ii)} A closed-form
solution for TDOA denoising for Gaussian noise and the presence of
missing data. \emph{iii)} Novel robust-denoising methods for handling
additive correlated noise, outliers and missing data.



\subsection{Notation}

Real scalar values are represented by lowercase letters (\eg
$\delta$). Vectors are by default arranged column-wise and are
represented by lowercase bold letters (\eg $\mbf{x}$). Matrices are
represented by uppercase bold letters (\eg $\TDOAMatrix$). Lower-case
letters are reserved to define vector and set sizes (\eg vector
$\mbf{x}=(x_1,\cdots,x_n)^\top$ is of size $n$), and $\mbf{x}^\top$
denotes transpose of vector $\mbf{x}$. Calligraphic fonts are reserved
to represent generic sets (e.g. $\mathcal{G}$) or functions applied to
matrices (e.g. $\mathcal{P}(\mbf{X})$). The $l_2$ norm $\|\cdot\|_2$
will be written by default as $\|\cdot\|$ for simplicity, and
$\|\cdot\|_F$ is the Frobenius norm, while $|\cdot|$ is reserved to
represent absolute values of scalars. The $l_0$ norm of a matrix, written
$\|\cdot\|_0$, is defined as the number of non-zero
elements of the matrix. $\mbf{A}\hp\mbf{B}$ is the Hadamard product between
$\mbf{A}$ and $\mbf{B}$, defined as the entrywise multiplication of
the corresponding matrices. $\tr(\cdot)$ is the trace function.

We also define the normalized unitary vector $\mbf{\hat{1}}$ as
$\mbf{\hat{1}}=\nobreak \lparen 1, \dotsc, 1 \rparen^\top/\sqrt{n}$, and
the null vector $\mbf{\hat{0}}$ as $\mbf{\hat{0}}=\nobreak \lparen 0,
\dotsc, 0 \rparen^\top$, both of them having size $n$. Finally,
$\mathds{1}=n~\mbf{\hat{1}}\mm\mbf{\hat{1}}^\top$ is a $n\times n$
matrix with all elements equal to $1$, $\mbf{D_x}$ is a $n\times n$
diagonal matrix where its main diagonal is the vector $\mbf{x}$, and
$\mbf{I}$ is the identity matrix.

\subsection{Paper Structure}

The rest of the paper is distributed as
follows. Section~\ref{sec:related-work} describes the related work and
Section~\ref{sec:problem-statement} the problem statement. In
section~\ref{sec:DefTDOA} TDOA matrices are described along with a
derivation of their properties. TDOA denoising in Gaussian noise
case is addressed in section~\ref{sec:denoising}, also providing a
closed-form solution. In sections~\ref{sec:robust-denoising}
and~\ref{sec:missing-data} we propose novel algorithms for robust
handling of noise and missing data,
respectively. Section~\ref{sec:robust-missing} combine the proposals of
the previous two sections into an unified algorithm. We also provide an
extensive experimentation to validate the proposed algorithms using both
synthetic (Section~\ref{sec:synthetic-data}) and real data
(Section~\ref{sec:real-data}). Finally, conclusions are drawn in
section~\ref{sec:conclusions}.

\section{Related Work}
\label{sec:related-work}

TDOA estimation is an essential first step for multiple applications
related to \textit{1)} localization, \textit{2)} self-calibration and \textit{3)} beamforming (among
others):

\textit{1) Localization}: widely used in radar, sonar and acoustics,
since no synchronization between the source and sensor is needed. The
TDOA information is combined with knowledge of the sensors' positions to
generate a Maximum Likelihood spatial estimator made from hyperboloids
intersected in some optimal sense. A linear closed-form solution of the
former problem, valid when the TDOA estimation errors are small, is
given in \cite{chan1994simple}.

\textit{2) Self-calibration}: since knowing the position of sensors is mandatory for localization
techniques, some strategies have been also proposed in order to
calibrate them using only TDOA measurements. In
\cite{pollefeys2008direct, kuang2013stratified}, the TDOA problem
is converted in a Time of Arrival (TOA) problem estimating the departure
time of signals. Then, self-calibration techniques for TOA can be
employed. The main drawback of this approach is that the conversion step
from TDOA to TOA is very sensitive to outliers and correlated noise.

\textit{3) Beamforming}: precise TDOA estimations is also critical for beamforming
  techniques and its applications. In \cite{anguera2007acoustic}, for
  example, additional steps are proposed for selecting the appropriate
  TDOA value among the correlation peaks, and also dealing with TDOA
  outliers. These steps include a Viterbi decoding based algorithm which
  maximizes the continuity of the TDOA estimations in several
  frames. However, the TDOA selection criteria is just based on their
  distance to surrounding TDOA values and their GCC-PHAT values, thus
  not attempting to benefit from the actual redundancy of the TDOA
  measurements.

Hence, an accurate estimation of TDOA is essential for a good
performance of any of the former applications based on these
measurements. 

Typically, when only two sensors are employed, the peak of the
generalized cross-correlation (GCC) function of the signals of two
sensors is a good estimator for the TDOA, for reasonable noise and
reverberation levels~\cite{dibiase2001robust}. 

When more than two sensors are used ($n>2$), there are $n(n-1)/2$
different TDOA measurements from all possible pairs of sensors, forming
the \emph{full TDOA set} or \emph{spherical
  set}~\cite{yang2006theoretical}.  However, all those TDOA measurements
are redundant. In fact, usually one sensor is considered the reference
sensor, and only the subset of $n-1$ TDOA measurements which involve
that sensor are considered. That \emph{non-redundant set} is the
set of measurement used by the majority of TDOA-based positioning
algorithms proposed in the literature~\cite{smith1987closed,
  chan1994simple, gillette2008linear, weng2011total, lin2013new,
  jamali2013sparsity}. Nevertheless, an optimal (denoised) version of
the non-redundant set can be estimated from the redundant set using a
Bayesian Linear Unbiased Estimator (BLUE), also known as the
Gauss-Markov estimator \cite{hahn73optimun}.

A closed-form solution for the BLUE estimator is provided in
\cite{so2008closed}, also proving that it is equal to the standard least
squares estimator, and that it reaches the Cramer-Rao lower bound for
positioning estimation. However, all the results in that work are based
on the assumption of additive Gaussian noise, which is unrealistic in
many practical applications~\cite{renaux2007unconditional}, and doesn't
yield good results when correlated noise is present as a consequence,
for instance, of multipath propagation. Additionally, the experimental
results shown in their work are only applied to synthetic data, thus not
allowing to assess the performance of their proposal in real scenarios
(in section~\ref{sec:real-data} we show the limitations of their method
when evaluated on real data). 

A least-squares solution to TDOA  denoising is proposed in \cite{schmidt1996least}.
 It is based on projecting the \emph{non-redundant set} of TDOA measurements into 
a set of ``feasible'' bivectors (rank 2, antisymmetric tensors) that show the
 same geometric properties of TDOA matrices. This denoising is also optimal for Gaussian 
noise but as in~\cite{renaux2007unconditional} the experimental analysis is based on 
simulated data and it does not cope with missing data or the presence of outliers 
in the TDOA measurements.

Periodicity in correlated signals, coherent noise and multi-path
due to reverberation are the major sources of non-Gaussian error in TDOA
estimation. Different approaches have been proposed to deal with them. A
basic method consists in making the GCC function more robust,
de-emphasizing the frequency-dependent weighting. The Phase Transform
(PHAT)~\cite{knapp1976GCC} is one example of this procedure which has
received considerable attention as the basis of acoustic source
localization systems due to its robustness in real world
scenarios~\cite{Zhang2008PHAT, velasco2014}. Other approaches are based
in blind estimation of multi-path (room impulse response)
\cite{benesty2000adaptive} but they need a good initialization to
perform well. 

Some previous works have also proposed more complicated structures in
order to represent TDOA redundancy, while not imposing strong
assumptions on the noise
distribution. In~\cite{scheuing2008disambiguation} a representation
based on graphs allows to disambiguate if a peak in correlation was
generated by the direct path or by reverberation applying an efficient
search algorithm among all possible combinations. However, they do not
explicitly attempt to provide improved TDOA estimations by exploiting
their redundancy.


Also different matrix representations have been used in the bibliography
regarding TDOA formulation. For example \cite{annibale2013tdoa} uses a
representation slightly different to the TDOA matrices we describe here,
but such representation does not have the algebraic properties that TDOA
matrices have, and their authors do not address an study in this sense.

So, to the best of our knowledge, there are no previous reported work
dealing with improving TDOA estimations by exploiting their redundancy,
while not imposing Gaussian noise restrictions, not requiring the sensor
positions, and being able to deal with the presence of outliers and
missing measurements (errors that will severely impact the performance
of applications based on TDOA measurements). In this paper we show that
TDOA matrices are a powerful tool that combined with recent advances in
robust low-rank estimation, are able to generate novel solutions for
these problems.







\section{Problem Statement}
\label{sec:problem-statement}

Hereafter, we assume only one source located at the position
$\mbf{r}=(r_x,r_y,r_z)^\top$, and $\numMics$ sensors synchronized
between them and placed in different positions
${\mbf{s}_i =\nobreak (s_{ix}, s_{iy},
  s_{iz})^\top},\,i\in\nobreak[1,\numMics]$.

Given this setup, we will assume that the source is emitting an unknown
signal $x(t)$. Then, the signal received by the sensor~$i$, $x_i(t)$, is
without loss of generality, a delayed and attenuated version of $x(t)$
(direct propagation) in addition to a signal $g_i(t)$ which summarizes
all the adverse effects, i.e. noise, interference, multipath, etc. Thus,
$x_i(t)=\nobreak x(t-\nobreak \tau_i)+\nobreak g_i(t)$, where $\tau_i=\|\mbf{r}-\mbf{s}_i\|_2/c$ is
the time of arrival (TOA) of the signal $x(t)$ at the sensor
$\mbf{s}_i$, being $c$ the propagation speed.

Assuming that TOA cannot be estimated directly, the time delay of
arrival (TDOA) between the sensors $i$ and $j$ is estimated by
correlating the received signals $x_i(t)$ and $x_j(t)$ (typically
using the Generalized Cross-Correlation
GCC~\cite{knapp1976GCC}).

\section{TDOA matrices}
\label{sec:DefTDOA}

In this section we define TDOA matrices, and develop their main
properties. In a nutshell, given any TDOA matrix $\TDOAMatrix$, we show
that: \emph{i)} $\TDOAMatrix$ is rank $2$ (Theorem~\ref{th:rank}),
\emph{ii)}~$\TDOAMatrix$ can be decomposed as $\TDOAMatrix=\nobreak\lparen\mbf{x}\mm\mbf{\hat{1}}^\top-\mbf{\hat{1}}\mm\mbf{x}^\top\rparen$
with $\mbf{x}=\TDOAMatrix \mm \mbf{\hat{1}}$
(Lemma~\ref{lemma:subspace}) and \emph{iii)} the previous decomposition is
bijective (Theorem~\ref{th:isomorphism}).

These properties are the foundations of the denoising algorithms that we
present in sections~\ref{sec:denoising} and~\ref{sec:robust-denoising},
and the missing data recovery proposal described in
section~\ref{sec:missing-data}, plus their combination described in
section~\ref{sec:robust-missing}.

\subsection{Definition of TDOA matrices}
\label{sec:defin-tdoa-matr}

\begin{definition} A TDOA matrix $\TDOAMatrix$, is a
  $(n \times n)$ skew-symmetric matrix where the element
  $(i,j)$
  is the time difference of arrival (TDOA) between the signals arriving
  at sensor $i$ and sensor $j$:
  \begin{equation}
    \label{eq:M_definition}
    \TDOAMatrix =\left\{\Delta\tau_{ij}\right\}=\begin{pmatrix}
      0 & \Delta\tau_{12} & \cdots & \Delta\tau_{1n} \\
      \Delta\tau_{21} & 0 & \cdots & \Delta\tau_{2n}\\
      \vdots & \vdots & \ddots & \vdots  \\
      \Delta\tau_{n1} & \Delta\tau_{n2} & \cdots & 0
    \end{pmatrix}
  \end{equation}
  with
  \begin{equation}
    \label{eq:tau}
    \Delta\tau_{ij}=\nobreak\lparen\tau_i-\tau_j\rparen,
  \end{equation}
  where $\tau_i$ is the time of arrival of the signal $x(t)$ at the
  sensor $\mbf{s}_i$. 
\end{definition}

We will also express $\TDOAMatrix$ in terms of its columns as $\TDOAMatrix = \nobreak \lparen \mbf{m}_1, \mbf{m}_2,
    \cdots , \mbf{m}_n \rparen$, being $\mbf{m}_i = \nobreak \lparen
\Delta\tau_{1i}, \Delta\tau_{2i}, \ldots, \Delta\tau_{ni}
\rparen^\top$.

We denote as $\TDOASETn$ to the set of TDOA matrices of size $n \times n$. 

Notice that there is a bijection between the full TDOA set and the
corresponding TDOA matrix. Nevertheless expressing TDOA measurements
as a matrix has important advantages, that we will discover throughout
this article. 

Note also that in the former definition, knowing the sensor array
geometry is not required. For a given geometry, all the feasible TDOA
matrices (those that are consistent with that particular geometry) are
a subset of $\TDOASETn$. Studying the properties of such subset is
out of the scope of this paper and the interested reader can refer
to~\cite{alameda2014geometric,compagnoni2014comprehensive,compagnoni2015denoising}
for further details.

\begin{figure}[!t]
  \centering
  \includegraphics[width=0.48\textwidth]{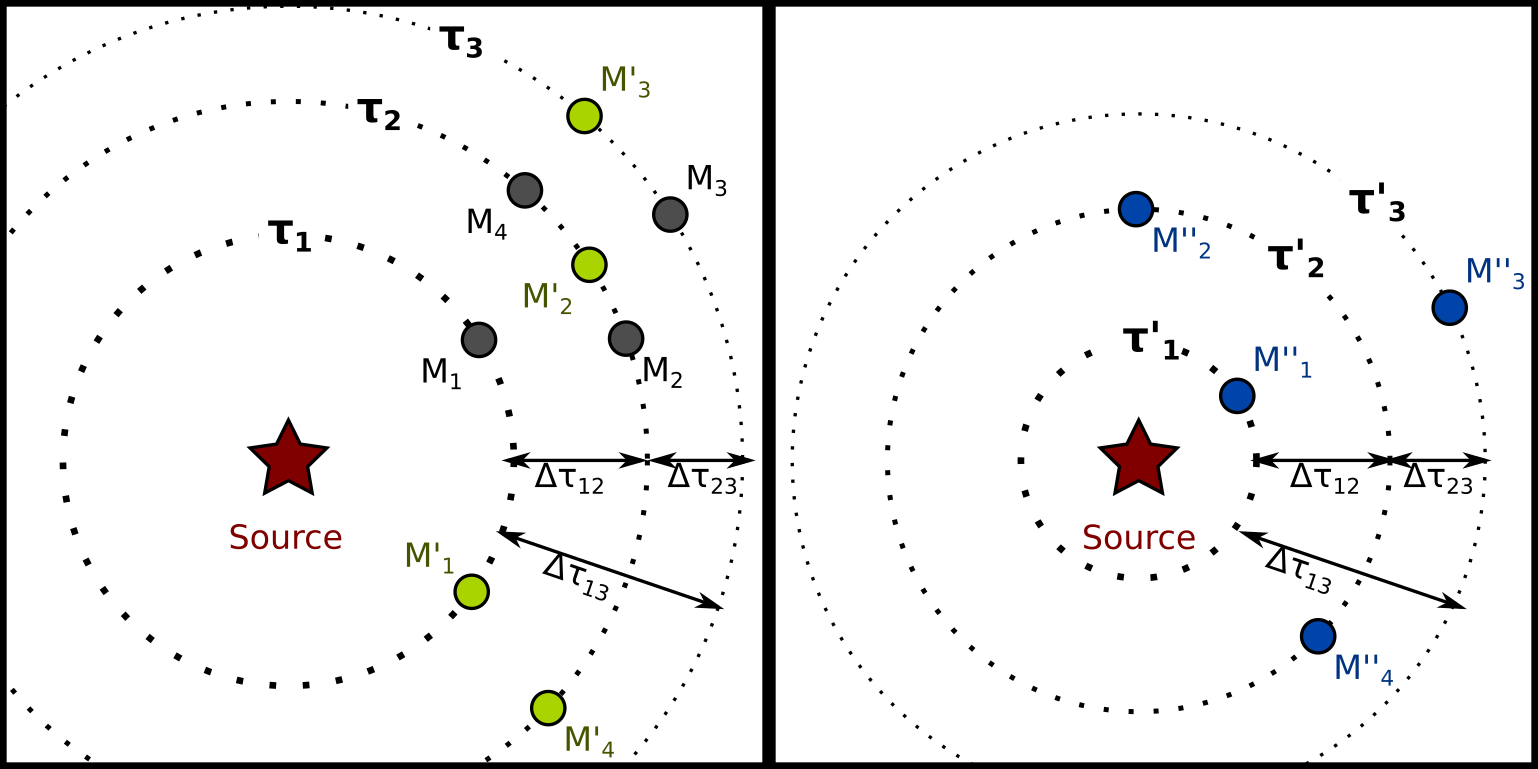}
  \caption{Example of three different geometrical configurations (grey, green and blue) of 4
    sensor with identical TDOA matrix.}
  \label{fig:same_tdoa}
\end{figure}

 On the other hand, given a particular TDOA matrix, there are infinite
number of sensors geometries which match with it. Left side of Fig.~\ref{fig:same_tdoa} 
shows that, given a set of TOAs ($\tau_1$,..., $\tau_n$) compatible with the set of TDOA measurements, the microphones can be situated in any place along the circumference
(sphere in the 3D case) with center in the source (dotted
lines), preserving its corespondent TOA (and therefore, its TDOA). Right side of Fig.~\ref{fig:same_tdoa} shows that there are a infinite number of TOA sets that comply with a given set of TDOA measurements.

\subsection{Rank of TDOA matrices}
\label{sec:rank_TDOA}

\begin{theorem}
  \label{th:rank}
  Let $\TDOAMatrix \in \TDOASETn$, then $\TDOAMatrix$ is rank 2 or 0
  (trivial case).
\end{theorem}
\begin{IEEEproof}
  The matrix $\TDOAMatrix$ can be expressed as:
  \begin{equation}
    \TDOAMatrix =\mbf{T}-\mbf{T}^\top,
  \end{equation}
  where $\mbf{T}$ is a rank 1 matrix defined as:
  \begin{equation}
    \label{eq:T_definition}
    \mbf{T}=\begin{pmatrix}
      \tau_1 & \cdots & \tau_1 \\
      \vdots  & \ddots & \vdots  \\
      \tau_n & \cdots & \tau_n
    \end{pmatrix},
  \end{equation}
  Applying the well known inequality:
  \begin{equation}
    \label{eq:rank_ineq}
    \rank(\mbf{A}+\mbf{B})\leq\rank(\mbf{A})+\rank(\mbf{B}),
  \end{equation}
  we can deduce that $\rank(\TDOAMatrix)\leq2$.  

  Moreover, since the rank of any skew-symmetric matrix must be even,
  rank 1 is not feasible. So we can conclude that, excepting the case
  that $\TDOAMatrix$ is the zero matrix (trivial case), the rank of
  $\TDOAMatrix$ is 2. This completes the proof.
\end{IEEEproof}

Note that the formerly referred as 'trivial case' only occurs when the
time of arrival is the same for all the sensors, i.e sensors are
geometrically placed on a sphere with center in the source.

Rank deficiency of TDOA matrices means that their rows and columns are
linearly dependent. That is consistent with the fact that, in the
noise-free case, the full TDOA set can be generated from the
non-redundant set using linear equations \cite{so2008closed}. In fact,
in a TDOA matrix, the column $j$ is the TDOA non-redundant set when
the sensor $j$ is the reference for TDOA measurements. Hereafter, and
without loss of generality, we will consider the first sensor as the
reference for the non redundant-set.


\subsection{Bijective mapping for TDOA matrices}
\label{sec:bijective-map}
We show next a particular representation of TDOA matrices with $n-1$ parameters that describes their spectral properties and also forms the base for our denoising algorithms.
\begin{theorem}
  \label{th:isomorphism}
  Let $H\subset\mathbb{R}^n$ be a $n-1$ dimensional hyperplane of
  $\mathbb{R}^n$ such that $\mrm{span}(\hat{1})\nsubseteq H$, then
  there always exits an isomorphism between $\TDOASETn$ and $H$ of form:
\begin{align*}
    \phi_H:H&\longrightarrow\quad\TDOASETn\\
    \mbf{z}&\longmapsto\mbf{z}\mm\mbf{\hat{1}}^\top-\mbf{\hat{1}}\mm\mbf{z}^\top
  \end{align*}

\end{theorem}
\begin{IEEEproof}
Theorem~\ref{th:isomorphism} states that a
  bijective linear map exists between $\TDOASETn$ an a $n-1$ dimensional
  subset of $\mathbb{R}^n$.
  Since the matrix $\mbf{T}$ in~\eqref{eq:T_definition} can be
  rewritten as $\mbf{T}=\mbf{z}\mm\mbf{\hat{1}}^\top$, where
  $z=(\tau_1,\cdots,\tau_n)^\top$, we can define the following linear map:
  \begin{align*}
    \phi:\mathbb{R}^n&\longrightarrow\quad\TDOASETn\\
    \mbf{z}&\longmapsto\mbf{z}\mm\mbf{\hat{1}}^\top-\mbf{\hat{1}}\mm\mbf{z}^\top
  \end{align*}
that is clearly surjective but not injective. That is, any
vector $\mbf{z}'=\mbf{z}+\alpha\mm\mbf{\hat{1}}$ represents the same
TDOA matrix. Indeed, the kernel of $\phi$ is
the linear subspace of $\mathbb{R}^n$ generated
by $\mbf{\hat{1}}$. 

Since we are looking for an isomorphism, we will restrict the domain of
$\phi$ to ensure injectivity. It yields $\phi_H$, where
$H\subset\mathbb{R}^n$ is a hyperplane of $\mathbb{R}^n$ not
containing $\mrm{ker}(\phi)$.  Note that $\phi_H$ is bijective as
$H\subset\mathbb{R}^n$ keeps the surjectivity of $\phi$ and, since
$\mrm{ker}(\phi_H)=\{\mbf{0}\}$, the function is also injective.
\end{IEEEproof}

Hereafter, we will only consider the particular case of $\phi_H$ for
the hyperplane $H=\mrm{ker}(\phi)^{\perp}$. It yields the following
expression for any $M\in\TDOASETn$:
\begin{equation}
   \label{eq:TDOA_decom}
   \TDOAMatrix
   =\lparen\mbf{x}\mm\mbf{\hat{1}}^\top-\mbf{\hat{1}}\mm\mbf{x}^\top\rparen\quad,~~\,\mbf{x}\perp\mbf{\hat{1}}\quad,\mbf{x}\in\mathbb{R}^n
\end{equation}

Choosing $\mbf{x}$ perpendicular to $\mbf{\hat{1}}$ is very
convenient, since both it simplifies $\phi_H^{-1}$ as shown in the
corollary~\ref{cor:inverse_map}, and it also can be used for calculate
the singular
value decomposition (SVD) of any $M\in\TDOASETn$, as discussed in~\ref{sec:svd}. This
leads to a parametric representation of $\TDOAMatrix$ that has important
properties that we will exploit later for TDOA denoising.
\begin{coro}
\label{cor:inverse_map}
 Given any  $\TDOAMatrix
\in \TDOASETn$, the corresponding vector $\mbf{x}\in\mathbb{R}^n$,
perpendicular to $\mbf{1}$, can be calculated as:
\begin{equation}
  \label{eq:x_from_TDOA}
   \mbf{x}=\TDOAMatrix\mm\mbf{\hat{1}}.
\end{equation}
\end{coro}
\begin{coro}
\label{cor:alter_decomp}
$\TDOAMatrix\in\TDOASETn$ can also be expressed as:

\begin{equation}
  \label{eq:TDOA_alter_decomp} 
  \TDOAMatrix =\frac{1}{\sqrt{n}}\lparen\mbf{D_x}\mm\mathds{1}-\mathds{1}\mm\mbf{D_x}\rparen\quad,\,\mbf{\hat{1}}\perp\mbf{\hat{x}},
\end{equation}
since $\mbf{x}\mm\mbf{\hat{1}}^\top=\mbf{D_x}\mm\mathds{1}/\sqrt{n}$
and $\mbf{\hat{1}}\mm\nobreak\mbf{x}^\top=\nobreak\mathds{1}\mm\nobreak\mbf{D_x}/\sqrt{n}$,
\end{coro}

\subsubsection{Singular Value Decomposition}
\label{sec:svd}

Because $\TDOAMatrix \in \TDOASETn$ is a skew-symmetric matrix of rank 2, it has the following singular value decomposition (SVD)~\cite[Supplementary
material]{ye2012robust}:
\begin{equation}
  \label{eq:TDOA_svd}
  \TDOAMatrix =\lparen\mbf{\hat{u}}_2,
    -\mbf{\hat{u}}_1\rparen \begin{pmatrix}\sigma&0\\0&\sigma\end{pmatrix}\lparen\mbf{\hat{u}}_1,
    \mbf{\hat{u}}_2\rparen^\top=\sigma  \lparen\mbf{\hat{u}}_2,
    -\mbf{\hat{u}}_1\rparen \lparen\mbf{\hat{u}}_1, \mbf{\hat{u}}_2\rparen^\top,
\end{equation}
where $\mbf{\hat{u}}_1$ and $\mbf{\hat{u}}_2$ are orthonormal vectors and $\sigma\geq 0$. Note that the SVD decomposition of $\mathbf{M}$ is not unique. Given any orthogonal $2\times2$  matrix $\mbf{R}$, the vectors $\lparen\mbf{\hat{v}}_1,\mbf{\hat{v}}_2\rparen=\lparen\mbf{\hat{u}}_1,\mbf{\hat{u}}_2\rparen\mm\mbf{R}$  also represent a valid SVD decomposition:
\begin{equation}
  \label{eq:TDOA_svdunique}
  \TDOAMatrix =\sigma\lparen\mbf{\hat{v}}_2,-\mbf{\hat{v}}_1\rparen\lparen\mbf{\hat{v}}_1,\mbf{\hat{v}}_2\rparen^\top=\sigma\lparen\mbf{\hat{u}}_2,-\mbf{\hat{u}}_1\rparen\lparen\mbf{\hat{u}}_1,\mbf{\hat{u}}_2\rparen^\top.
\end{equation}
Among all possible SVD decompositions of $\TDOAMatrix$ note that
theorem~\ref{th:isomorphism} guarantees that always exists one where
$\mbf{\hat{u}}_1=\mbf{\hat{1}}$. In such case SVD decomposition
can be computed from~\eqref{eq:TDOA_decom} as stated in the following theorem:
\begin{lemma}
  \label{lemma:subspace}

  Given $\TDOAMatrix \in \TDOASETn$, it admits the following SVD decomposition: 
  \begin{equation}
    \label{eq:decomp}
    \mathbf{M}=\sigma\lparen\mbf{\hat{u}},-\mbf{\hat{1}}\rparen\lparen\mbf{\hat{1}},\mbf{\hat{u}}\rparen^\top \quad \text{with} \quad \mbf{\hat{u}}=\dfrac{\TDOAMatrix \mm \mbf{\hat{1}}}{\|\TDOAMatrix \mm \mbf{\hat{1}}\|} \quad \sigma=\|\TDOAMatrix \mm \mbf{\hat{1}}\|
  \end{equation}
\end{lemma}

Note that, according the previous lemma, not all rank 2
skew-symmetric matrices are TDOA-Matrices. In fact, if $\mbf{\hat{v}}_1$
and $\mbf{\hat{v}}_2$ in~\eqref{eq:TDOA_svdunique} are not coplanar
with $\mbf{\hat{1}}$
(i.e. $(\mbf{\hat{1}}^\top\mbf{\hat{v}}_1)\mbf{\hat{v}}_1+(\mbf{\hat{1}}^\top\mbf{\hat{v}}_2)\mbf{\hat{v}}_2\neq
\mbf{\hat{1}}$), then the resulting matrix is rank 2 and
skew-symmetric but not a TDOA matrix.

\section{TDOA Denoising}
\label{sec:denoising}

In this section we propose a denoising strategy to deal with Gaussian
noise in the estimated TDOA measurements, deriving a closed form
solution for the proposed optimization problem. This solution is also
compared with the Gauss-Markov Estimator.

\subsection{Denoising Strategy}
\label{sec:denoising-strategy}

We assume now that each TDOA measurement is contaminated with
uncorrelated Gaussian noise $n_{ij}=-n_{ji}$, such that
$\Delta\tilde{\tau}_{ij}=\Delta\tau_{ij}+n_{ij}$. Therefore, the
measured TDOA matrix
$\tilde{\TDOAMatrix}=\left\{\Delta\tilde{\tau}_{ij}\right\}$ 
is also a skew-symmetric matrix, sum of a noise-free $\TDOAMatrix \in \TDOASETn$ and a
skew-symmetric matrix containing noise $\mbf{N}=\left\{n_{ij} \right\}$:
\begin{equation}
  \label{eq:noisy_M}
  \tilde{\TDOAMatrix}=\TDOAMatrix +\mbf{N}.
\end{equation}
Because of the noise, $\tilde{\TDOAMatrix} \notin \TDOASETn$ and thus
Theorem~\ref{th:rank} is no longer satisfied. Consequently, the rank of
$\tilde{\TDOAMatrix}$ may be higher than two. Nevertheless, we will show
that we can take advantage of the structure of TDOA matrices in order to
denoise the measured data.

For denoising, we propose finding the closest $\TDOAMatrix^{\ast}
\in\nobreak \TDOASETn$, to the measured matrix $\tilde{\TDOAMatrix}$, in
the sense of the Frobenius norm. This approach yields the following
optimization problem:
\begin{equation}
  \label{eq:denoising}
  \TDOAMatrix^{\ast}= \underset{\TDOAMatrix \in \TDOASETn
  }{\operatorname{arg\,min}}\quad\left\|\tilde{\TDOAMatrix}-\TDOAMatrix \right\|_F^2.
\end{equation}

\subsection{Closed-Form Solution}
\label{sec:closed-form}
\begin{theorem}
  \label{th:closed-form}
  Problem~\eqref{eq:denoising} has the following closed form solution: $ \TDOAMatrix^\ast=(\tilde{\TDOAMatrix}\mm\mathds{1}+\mathds{1}\mm\tilde{\TDOAMatrix})/n$
\end{theorem}
\begin{IEEEproof}
  From \eqref{eq:TDOA_decom}, the denoising problem
  \eqref{eq:denoising} is equivalent to the following constrained
  convex optimization problem:
  \begin{equation}
    \label{eq:fobenius-min}
    \begin{aligned}
      & \underset{\mbf{x}}{\operatorname{minimize}} & &
      \left\|\tilde{\TDOAMatrix}-\lparen\mbf{x}\mm\mbf{\hat{1}}^\top-\mbf{\hat{1}}\mm\mbf{x}^\top\rparen\right\|_F^2 \\
      & \operatorname{subject\;to} & &  \mbf{\hat{1}}^\top\mm\mbf{x}=0.
    \end{aligned}
  \end{equation}
 Using the definition of Frobenius norm 
  $\|\mbf{A}\|_F^2=\tr(\mbf{A} \mbf{A}^\top)$, and trace properties
  $\tr(\mbf{A} \mbf{B})=\tr(\mbf{B} \mbf{A})$ and
  $\tr(\mbf{A})=\tr(\mbf{A}^\top)$ we rewrite the cost as:

  \begin{multline}
    \label{eq:frob_equiv}
    \left\|\tilde{\TDOAMatrix}-\lparen\mbf{x}\mm\mbf{\hat{1}}^\top-\mbf{\hat{1}}\mm\mbf{x}^\top\rparen\right\|_F^2= \\
    =\tr\lparen\left[\tilde{\TDOAMatrix}-\lparen\mbf{x}\mm\mbf{\hat{1}}^\top-\mbf{\hat{1}}\mm\mbf{x}^\top\rparen\right]\left[\tilde{\TDOAMatrix}-\lparen\mbf{x}\mm\mbf{\hat{1}}^\top-\mbf{\hat{1}}\mm\mbf{x}^\top\rparen\right]^\top\rparen\\
    =2\lparen\mbf{x}^\top\mm\mbf{x} -\mbf{x}^\top\mm\mbf{\hat{1}}\mm\mbf{\hat{1}}^\top\mm\mbf{x}
      -\mbf{\hat{1}}^\top\mm\tilde{\TDOAMatrix}^\top\mm\mbf{x}
      +\mbf{\hat{1}}^\top\mm\tilde{\TDOAMatrix}\mm\mbf{x}
    \rparen+\\
    + \tr{\lparen\tilde{\TDOAMatrix}\mm\tilde{\TDOAMatrix}^\top\rparen=f\lparen\mbf{x};\tilde{\TDOAMatrix}\rparen}.
  \end{multline}

  To solve the constrained problem \eqref{eq:fobenius-min} we use the method of Lagrange multipliers, resulting in the following unconstrained equivalent: 
  \begin{equation}
    \label{eq:denoising_dualproblem}
    \mbf{x}^\ast\,=\,\underset{\mbf{x},\lambda}{\operatorname{arg\,min}}\,\left[\Lambda\lparen\mbf{x};\lambda\rparen\right],
  \end{equation}
  where $\lambda$ is the Lagrange multiplier and
  \begin{equation}
    \label{eq:def_optfun}
    \Lambda\lparen\mbf{x};\lambda\rparen=f\lparen\mbf{x};\tilde{\TDOAMatrix}\rparen+\lambda\mbf{\hat{1}}^\top\mm\mbf{x}.
  \end{equation}

 We find extrema in \eqref{eq:def_optfun} by taking first derivatives with respect to both $\mbf{x}$ and $\lambda$ and solving the following system:
 \begin{multline*}
   \nabla\Lambda\lparen\mbf{x};\lambda\rparen=\mbf{\hat{0}}\,\Rightarrow\\
\begin{cases} 4 \mbf{x}^\top\lparen \mbf{I}-\mbf{\hat{1}}\mm\mbf{\hat{1}}^\top\rparen+2\mbf{\hat{1}}^\top\lparen\tilde{\TDOAMatrix}-\tilde{\TDOAMatrix}^\top\rparen+\lambda\mbf{\hat{1}}^\top=\mbf{\hat{0}}^\top\\ \mbf{\hat{1}}^\top\mm\mbf{x}=0\\
\end{cases}
 \end{multline*}
\begin{align}
        \mbf{x}^\ast&=\frac{\lparen\tilde{\TDOAMatrix}-\tilde{\TDOAMatrix}^\top\rparen\mbf{\hat{1}}-\lambda\mbf{\hat{1}}}{2}\nonumber\\
        \lambda^\ast&=\frac{\mbf{\hat{1}}^\top\lparen\tilde{\TDOAMatrix}-\tilde{\TDOAMatrix}^\top\rparen\mbf{\hat{1}}}{2\mbf{\hat{1}}^\top\mm\mbf{\hat{1}}}.
\label{eq:pre_denoising}
\end{align}

Given that the objective function is strictly convex, the solution to
the system is unique, and therefore because the proposed solution
($\mbf{x}^\ast$ and $\lambda^\ast$) satisfies the equations of a
critical point, it is the global minimum

Since $\tilde{\TDOAMatrix}$ is skew-symmetric $(\tilde{\TDOAMatrix}-\tilde{\TDOAMatrix}^\top)=2\tilde{\TDOAMatrix}$. Therefore, \eqref{eq:pre_denoising} becomes:
\begin{subequations}
\begin{align}
        \mbf{x}^\ast&=\frac{2\tilde{\TDOAMatrix}\mm\mbf{\hat{1}}-\lambda\mbf{\hat{1}}}{2}=\tilde{\TDOAMatrix}\mm\mbf{\hat{1}}\label{eq:denoising_x}\\
        \lambda&=\frac{2\mbf{\hat{1}}^\top\mm\tilde{\TDOAMatrix}\mm\mbf{\hat{1}}}{2\mbf{\hat{1}}^\top\mm\mbf{\hat{1}}}=2\mbf{\hat{1}}^\top\mm\tilde{\TDOAMatrix}\mm\mbf{\hat{1}}=0.\label{eq:denoising_lambda}
\end{align}
\end{subequations}
In \eqref{eq:denoising_lambda} we use the fact that
$\mbf{\hat{1}}^\top\mm\mbf{A}\mm\mbf{\hat{1}}=0$ for $\mbf{A}$ being a
skew-symmetric matrix.  Also, it is interesting to note from
\eqref{eq:denoising_x} that $\mbf{x}^\ast$ follows the same expression
as the one stated in \eqref{eq:TDOA_decom} for $\mbf{x}$ in
the noise-free case.

A compact expression for $\TDOAMatrix^\ast$ can be easily derived from
\eqref{eq:denoising_x} via \eqref{eq:TDOA_decom}:
\begin{multline}
  \label{eq:closed-form}
  \TDOAMatrix^\ast=\lparen\mbf{\hat{1}},\tilde{\TDOAMatrix}\mm\mbf{\hat{1}}\rparen\lparen-\tilde{\TDOAMatrix}\mm\mbf{\hat{1}},\mbf{\hat{1}}\rparen^\top=\tilde{\TDOAMatrix}\mm\mbf{\hat{1}}\mm\mbf{\hat{1}}^\top+\mbf{\hat{1}}\mm\mbf{\hat{1}}^\top\mm\tilde{\TDOAMatrix}=\\
=(\tilde{\TDOAMatrix}\mm\mathds{1}+\mathds{1}\mm\tilde{\TDOAMatrix})/n.
\end{multline}
This completes the proof.
\end{IEEEproof}
 
Since lemma~\ref{lemma:subspace} relates \eqref{eq:TDOA_decom} with
SVD, the proposed denoising approach can be considered as a version of
Eckart-Young-Mirsky theorem \cite{eckart1936approximation,mirsky1960symmetric} constrained to TDOA matrices.



\subsection{Equivalence with the Gauss-Markov Estimator} 
\label{sec:equiv-GaussMarkov}

By operating in \eqref{eq:closed-form}, each element $(i,j)$ of the denoised matrix $\TDOAMatrix^\ast$ is obtained as follows:
\begin{equation}
  \label{eq:denoised-element}
  \TDOAMatrix^\ast=\{\Delta\tau_{ij}^\ast\}=\left\{\frac{1}{n}\lparen\sum_{k=1}^n{\Delta\tau_{ik}+\Delta\tau_{kj}}\rparen\right\}.
\end{equation}

The closed-form in \eqref{eq:denoised-element} is identical to the one
reported in \cite[eq.(14)]{so2008closed} as the Gauss-Markov estimator of
the TDOA measurements, so that all the properties there can be
extrapolated to this work. This is not surprising as the least-squares cost of \eqref{eq:denoising} is optimal for Gaussian noise. The same denoising result was found in \cite{schmidt1996least} by projecting TDOA measurements into the set of ``feasible'' bivectors in a least-squares sense. Under the assumption of Gaussian noise, we can conclude that \cite{so2008closed,schmidt1996least} and our denoising result in \eqref{eq:denoised-element} are completely equivalent. 


\section{Robust TDOA Denoising}
\label{sec:robust-denoising}

In some application scenarios, the assumption of uncorrelated white
noise made in section~\ref{sec:denoising} is fully unrealistic.
In cases where the noise is correlated with the signal, measurements are
prone to contain outliers in the TDOA measurements due to spurious
peaks in the correlation. For such cases, a more complete model for the
measured matrix is:
\begin{equation}
  \label{eq:nongaussian_M}
  \tilde{\TDOAMatrix}=\TDOAMatrix+\mbf{N}+\mbf{S},
\end{equation}
where $\TDOAMatrix \in \TDOASETn$, $\mbf{N}$ is a skew-symmetric matrix
containing Gaussian noise, much like in \eqref{eq:noisy_M}, and the new
matrix $\mbf{S}$ models the addition of all the outliers. Since the
number of outliers is usually small as compared with the number of
measurements, we will assume $\mbf{S}$ to be sparse and unknown.

In order to denoise $\tilde{\TDOAMatrix}$, we propose solving the
following optimization problem, finding both matrices $\TDOAMatrix$ and
$\mbf{S}$:
\begin{equation}
  \label{eq:robust-denoising}
  \begin{aligned}
    & \underset{\TDOAMatrix,\mbf{S}}{\operatorname{minimize}} & &
    \left\|\tilde{\TDOAMatrix}-\TDOAMatrix-\mbf{S}\right\|_F^2 \\
    & \operatorname{subject\;to} & & \TDOAMatrix \in \TDOASETn\\
    & & & \|\mbf{S}\|_0 < 2k,
  \end{aligned}
\end{equation}
where $k$ is the maximum number of outliers supposed to be present in
the TDOA measurements.




Robust denoising in \eqref{eq:robust-denoising} is a non-convex optimization problem with constraints that are not even differentiable. This kind of optimization problems have been explored in Robust PCA (RPCA)~\cite{candes2011robust} or robust low-rank factorizations such in GoDec~\cite{zhou2011godec}. Despite TDOA matrices are low-rank, these algorithms are not well suited here as they do not include all the algebraic constraints in TDOA matrices. 

In order to solve \eqref{eq:robust-denoising}, we propose an iterative
algorithm, inspired in GoDec. It consists of an alternation method in
which $\TDOAMatrix$ and $\mbf{S}$ are obtained in turns, with close-form
solutions for these two steps (we use a subindex $t$ to denote the
iteration count):
\begin{equation}
  \label{eq:modified_godec}
  \left\{
    \begin{aligned}
      \TDOAMatrix_t &= \underset{\TDOAMatrix \in \TDOASETn}{\operatorname{arg\,min}}\quad
      \left\|\tilde{\TDOAMatrix}-\TDOAMatrix-\mbf{S}_{t-1}\right\|_F^2\\
      \mbf{S}_t &= \underset{\|\mbf{S}\|_0<2k}{\operatorname{arg\,min}}\quad
      \left\|\tilde{\TDOAMatrix}-\TDOAMatrix_t-\mbf{S}\right\|_F^2\\
    \end{aligned}\right.
\end{equation}

The first sub-problem of \eqref{eq:modified_godec} is the same as our
denoising problem in \eqref{eq:denoising}, therefore $\TDOAMatrix_t$ can be updated via
\eqref{eq:closed-form}. Then, $\mbf{S}_t$ is updated via entry-wise
hard thresholding of $\tilde{\TDOAMatrix}-\TDOAMatrix_t$. Thus:

\begin{equation}
  \label{eq:modified_godec2}
  \left\{
    \begin{aligned}
      \TDOAMatrix_t &= \lparen\tilde{\TDOAMatrix}-\mbf{S}_{t-1}\rparen\,\mbf{\hat{1}}\mbf{\hat{1}}^\top+\mbf{\hat{1}}\mbf{\hat{1}}^\top\lparen\tilde{\TDOAMatrix}-\mbf{S}_{t-1}\rparen\\
      \mbf{S}_t &= \mathcal{P}_{2k}\lparen\tilde{\TDOAMatrix}-\TDOAMatrix_t\rparen
    \end{aligned}\right.
\end{equation}
where $\mathcal{P}_{l}(\mbf{X})$ is an function which generates a
matrix with the same size of $\mbf{X}$, preserving the $l$ elements of
$\mbf{X}$ with the largest absolute value, and making the rest of
elements zero. Note that, since $\mbf{X}$ is skew symmetric in our
application, the result provided by $\mathcal{P}_{2k}(\cdot)$ is also
skew symmetric. 
The convergence to a local minimum of this algorithm is guaranteed in similar circumstances as GoDec~\cite{zhou2011godec}, as the solutions to both sub-problems in \eqref{eq:modified_godec2} are solved globally. 

So, the proposed robust denoising algorithm is shown in Alg.~\ref{alg:roden}.


  \begin{algorithm}[H]
  \caption{Robust denoising.}\label{alg:roden}
  \textbf{Require:} $\tilde{\TDOAMatrix}$, $k$, $\epsilon$ 
  
  \textbf{Ensure:} $\TDOAMatrix\in\TDOASETn$, $\|\mbf{S}\|_0<2k$, 
  
  $\enspace$\footnotesize{1:}$\enspace$\normalsize $\TDOAMatrix_0=\tilde{\TDOAMatrix}$ ; $\mbf{S}_0=0$ ; $t=0$
  
  $\enspace$\footnotesize{2:}$\enspace$\normalsize \textbf{while} $\|\tilde{\TDOAMatrix}-\TDOAMatrix_t-\mbf{S}_t\|_F^2/\|\tilde{\TDOAMatrix}\|_F^2>\epsilon$ \textbf{do}
  
  $\enspace$\footnotesize{3:}$\enspace$\normalsize $\quad$ $t=t+1$
  
  $\enspace$\footnotesize{4:}$\enspace$\normalsize $\quad$ $\TDOAMatrix_t=(\tilde{\TDOAMatrix}-\mbf{S}_{t-1})\mbf{\hat{1}}\mbf{\hat{1}}^\top+\mbf{\hat{1}}\mbf{\hat{1}}^\top(\tilde{\TDOAMatrix}-\mbf{S}_{t-1})$
  
  $\enspace$\footnotesize{5:}$\enspace$\normalsize $\quad$ $\mbf{S}_t=\mathcal{P}_{2k}(\tilde{\TDOAMatrix}-\TDOAMatrix_t)$ 
  
  $\enspace$\footnotesize{6:}$\enspace$\normalsize \textbf{end while}
  
  $\enspace$\footnotesize{7:}$\enspace$\normalsize \textbf{return} $\TDOAMatrix_t$, $\mbf{S}_t$
\end{algorithm}


\ifthenelse{\equal{\useAlgNames}{true}}
{
From now on, we will refer to this algorithm as \RobustDenoisingAlg.
}


\section{Missing Data Recovery}
\label{sec:missing-data}

\subsection{Recovery Strategy}
\label{sec:recovery-strategy}

In real scenarios, there may be situations where some of the elements
of $\mbf{\tilde{M}}$ might not be available (for instance, due to
communications failure) or even when they are available,
there are reasons to avoid using them (for example, due to a priori
knowledge of unreliable measurements, or when calculating the whole
redundant set is computationally too
demanding)~\cite{compagnoni2015denoising}.
In such cases, we want to be able to avoid some
measurements, thus performing estimations when part of the values in
$\mbf{\tilde{M}}$ are missing. 

In this section, we address the matrix completion
problem (\cite{goldberg2010,candes2010}) for TDOA matrices. We
assume that in a measured TDOA matrix $\mbf{\tilde{M}}$, some of
its elements are unknown, and the rest are contaminated with additive
Gaussian noise. We take advantage of the redundancy present in TDOA
matrices to estimate a complete denoised TDOA matrix including the
missing entries.



The matrix completion problem is stated as follows:
\begin{equation}
    \label{eq:matrix_completion}
   \TDOAMatrix^{\ast}= \underset{\TDOAMatrix \in \TDOASETn}{\operatorname{arg\,min}}\quad\left\|\mbf{L}\hp\lparen\tilde{\TDOAMatrix}-\TDOAMatrix\rparen\right\|_F^2,
  \end{equation}
  where $\mbf{L}$ is a symmetric binary matrix whose element $(i,j)$ is
  $1$ if the TDOA between the sensor $i$ and $j$ is known, being $0$
  otherwise. For convenience and without loss of generality, the
  elements on the main diagonal of $\mbf{L}$ will be set to $1$.

  Solving \eqref{eq:matrix_completion} is equivalent to finding the full
  TDOA matrix whose elements best fit the available elements of
  $\mbf{\tilde{M}}$.  Note that,
  $\mbf{L}\hp(\tilde{\TDOAMatrix}-\TDOAMatrix)=(\tilde{\TDOAMatrix}_{\mbf{L}}-\mbf{L}\hp\TDOAMatrix)$,
  where
  $\tilde{\TDOAMatrix}_{\mbf{L}}=(\mbf{L}\hp\tilde{\TDOAMatrix})$ is
  the result of setting the unknown elements of $\mbf{\tilde{M}}$ to
  zero.

\subsection{Closed-Form Solution}
\label{sec:closed-form-completion}

\begin{theorem}
  \label{th:closed-form-completion}
  The problem~\eqref{eq:matrix_completion} has the following closed form solution: $ \TDOAMatrix^\ast=\lparen\mbf{D_\beta}+\mbf{\bar{L}}\rparen^{-1}\tilde{\TDOAMatrix}_{\mbf{L}}\,\mathds{1}+\mathds{1}\tilde{\TDOAMatrix}_{\mbf{L}}\lparen\mbf{D_\beta}+\mbf{\bar{L}}\rparen^{-1}$
where $\mbf{D_\beta}=\lparen\mbf{I}\hp\mbf{L}\mbf{L}^\top\rparen$ is a $n\times n$
diagonal matrix with $\pmb{\beta}=\lparen n-\bar{\beta}_1,\cdots,n-\bar{\beta}_n\rparen^\top=\sqrt{n}\mm\mbf{L}\mm\mbf{\hat{1}}$ as its
main diagonal. $\bar{\beta}_i$ is the number of missing measurements
with the sensor $i$.
\end{theorem}
\begin{IEEEproof}
 Using Corollary \ref{cor:alter_decomp}, problem \eqref{eq:matrix_completion} is rewritten as
  \begin{equation}
    \label{eq:completion_equiv}
        \begin{aligned}
      & \underset{\mbf{x}}{\operatorname{minimize}} & &
      \left\|\tilde{\TDOAMatrix}_\mbf{L}-\frac{\mbf{L}\hp\lparen\mbf{D_x}\mathds{1}-\mathds{1}\mbf{D_x}\rparen}{\sqrt{n}}\right\|_F^2 \\
      & \operatorname{subject\;to} & &  \mbf{\hat{1}}^\top\mm\mbf{x}=0.
    \end{aligned}
  \end{equation}
 Since $\mathds{1}$ is the identity element of the hadamard product and
  $\mbf{D_x}$ is a diagonal matrix, we can rewrite
  \eqref{eq:completion_equiv} as:
  \begin{equation}
    \label{eq:completion_equiv2}
        \begin{aligned}
      & \underset{\mbf{x}}{\operatorname{minimize}} & &
      \left\|\tilde{\TDOAMatrix}_\mbf{L}-\frac{\lparen\mbf{D_x}\mbf{L}-\mbf{L}\mbf{D_x}\rparen}{\sqrt{n}}\right\|_F^2 \\
      & \operatorname{subject\;to} & &  \mbf{\hat{1}}^\top\mm\mbf{x}=0.
    \end{aligned}
  \end{equation}
Operating in a similar manner to \eqref{eq:frob_equiv} we get:
\begin{multline}
  \label{eq:completion_trace}
  \left\|\tilde{\TDOAMatrix}_\mbf{L}-\frac{\lparen\mbf{D_x}\mbf{L}-\mbf{L}\mbf{D_x}\rparen}{\sqrt{n}}\right\|_F^2=\frac{2}{n}\tr\lparen\mbf{D_x}\mbf{L}\mbf{L}^\top\mbf{D_x}\rparen-\\
  -
  \frac{2}{n}\tr\lparen\mbf{D_x}\mbf{L}\mbf{D_x}\mbf{L}\rparen+\frac{2}{\sqrt{n}}\tr\lparen\left[\tilde{\TDOAMatrix}_\mbf{L}-\tilde{\TDOAMatrix}_\mbf{L}^\top\right]\mbf{D_x}\mbf{L}\rparen+\\
  +\tr\lparen\tilde{\TDOAMatrix}_\mbf{L}\tilde{\TDOAMatrix}_\mbf{L}^\top\rparen.
\end{multline}
Using the identity
$\mbf{x}^\ast\lparen\mbf{A}\hp\mbf{B}\rparen\mbf{y}=\tr\lparen\mbf{D_x}^\ast\mbf{A}\mbf{D_y}\mbf{B}^\top\rparen$ we get:
\begin{multline}
  \label{eq:completion_prev_final}
  \left\|\tilde{\TDOAMatrix}_\mbf{L}-\frac{\lparen\mbf{D_x}\mbf{L}-\mbf{L}\mbf{D_x}\rparen}{\sqrt{n}}\right\|_F^2=\frac{2}{n}\mbf{x}^\top\mm\lparen\mbf{I}\hp\mbf{L}\mbf{L}^\top\rparen\mm\mbf{x}\\
  -\frac{2}{n}\mbf{x}^\top\mm\lparen\mbf{L}\hp\mbf{L}^\top\rparen\mm\mbf{x} +2\mm\mbf{\hat{1}}^\top\mm\lparen\left[\tilde{\TDOAMatrix}_\mbf{L}-\tilde{\TDOAMatrix}_\mbf{L}^\top\right]\hp \mbf{L}^\top\rparen\mm\mbf{x} +\\
 + \tr\lparen\tilde{\TDOAMatrix}_\mbf{L}\tilde{\TDOAMatrix}_\mbf{L}^\top\rparen =g\lparen\mbf{x};\tilde{\TDOAMatrix},\,\mbf{L}\rparen
\end{multline}
and finally:
\begin{multline}
  \label{eq:completion_final}
  \left\|\tilde{\TDOAMatrix}_\mbf{L}-\frac{\lparen\mbf{D_x}\mbf{L}-\mbf{L}\mbf{D_x}\rparen}{\sqrt{n}}\right\|_F^2=\\
\frac{2}{n}\lparen\mbf{x}^\top\mm\mbf{D_\beta}\mm\mbf{x}-\mbf{x}^\top\mm\mbf{L}\mm\mbf{x} + n\mm\mbf{\hat{1}}^\top\mm\lparen\tilde{\TDOAMatrix}_\mbf{L}-\tilde{\TDOAMatrix}_\mbf{L}^\top\rparen\mm\mbf{x}\rparen + \\
+\tr\lparen\tilde{\TDOAMatrix}_\mbf{L}\tilde{\TDOAMatrix}_\mbf{L}^\top\rparen=g(\mbf{x};\tilde{\TDOAMatrix},\,\mbf{L}).
\end{multline}

It is important to note that equations \eqref{eq:frob_equiv} and
\eqref{eq:completion_final} are identical when there is no missing data in $\tilde{\TDOAMatrix}$ (i.e
$\mbf{L}=\mathds{1}=n\mm\mbf{\hat{1}}\mm\mbf{\hat{1}}^\top$ and
$\mbf{D_\beta}=n\mm\mbf{I}$). 

We use the method of Lagrange multipliers to express
\eqref{eq:completion_equiv} as the following unconstrained optimization
problem:

 
 \begin{equation}
    \label{eq:def_optfun2}
    \Lambda\lparen\mbf{x};\lambda\rparen=g\lparen\mbf{x};\tilde{\TDOAMatrix},\,\mbf{L}\rparen+\lambda\mbf{\hat{1}}^\top\mm\mbf{x}.
  \end{equation}
 By taking derivatives we obtain the following system:
 \begin{multline}
 \label{eq:gradient_completion}
   \nabla\Lambda\lparen\mbf{x};\lambda\rparen=\mbf{\hat{0}}\,\Rightarrow\\
\begin{cases}\frac{2}{n}\mbf{x}^\top\lparen\mbf{D_\beta}-\mbf{L}\rparen+\mbf{\hat{1}}^\top\lparen\tilde{\TDOAMatrix}_{\mbf{L}}-\tilde{\TDOAMatrix}_{\mbf{L}}^\top\rparen+\lambda\mbf{\hat{1}}^\top=\mbf{\hat{0}}\\
\mbf{\hat{1}}^\top\mbf{x}=0.
\end{cases}
 \end{multline}
 Since  $\mbf{\hat{1}}^\top\mbf{x}=0$ implies that $\mathds{1}\mbf{x}=\mbf{\hat{0}}$, we
 substitute in \eqref{eq:gradient_completion} obtaining:
\begin{equation}
  \label{eq:solution1_completion}
  \frac{2}{n}\lparen\mbf{D_\beta}+\mbf{\bar{L}}\rparen\mbf{x}=\lparen\tilde{\TDOAMatrix}_{\mbf{L}}-\tilde{\TDOAMatrix}_{\mbf{L}}^\top\rparen\mbf{\hat{1}}-\lambda\mbf{\hat{1}},
\end{equation}
where $\lparen\mbf{\bar{L}}=\mathds{1}-\mbf{L}\rparen$ (logical not operator over all elements of $\mbf{L}$). Note that: $(\mbf{D_\beta}+\mbf{\bar{L}})$ is symmetric and, furthermore, $\mbf{\hat{1}}$ is one of its eigenvectors.
\begin{equation}
  \label{eq:eigenvector}
  \lparen\mbf{D_\beta}+\mbf{\bar{L}}\rparen\mbf{\hat{1}}=\frac{\pmb{\beta}+\pmb{\bar{\beta}}}{\sqrt{n}}=n\mbf{\hat{1}}.
\end{equation}
Therefore, if the two terms of \eqref{eq:solution1_completion} are
multiplied on the right by $\mbf{\hat{1}}^\top$ we get:
\begin{equation}
  \label{eq:lambda_completion}
  2 \mm \mbf{\hat{1}}^\top\mbf{x}=\mbf{\hat{1}}^\top\lparen\tilde{\TDOAMatrix}_{\mbf{L}}-\tilde{\TDOAMatrix}_{\mbf{L}}^\top\rparen\mbf{\hat{1}}-\lambda.
\end{equation}
Then applying to \eqref{eq:lambda_completion} the fact that 
\begin{equation}
  \label{eq:tmp}
  \mbf{\hat{1}}^\top\mbf{x}=0 \quad \text{and} \quad \mbf{\hat{1}}^\top (\tilde{\TDOAMatrix}_{\mbf{L}}-\tilde{\TDOAMatrix}_{\mbf{L}}^\top) \mbf{\hat{1}}=0,
\end{equation}
we can conclude that $\lambda=0$. Thus, the solution of \eqref{eq:completion_equiv} is:

\begin{equation}
  \label{eq:solution_completion}
  \mbf{x}^{\ast}=n\lparen\mbf{D_\beta}+\mbf{\bar{L}}\rparen^{-1}\tilde{\TDOAMatrix}_{\mbf{L}}\mbf{\hat{1}}.
\end{equation}

Finally, the solution of problem \eqref{eq:matrix_completion} can be
calculated from \eqref{eq:solution_completion} using
\eqref{eq:TDOA_decom}:
\begin{multline}
  \label{eq:closed-form-completion}
  \TDOAMatrix^\ast=\lparen\mbf{D_\beta}+\mbf{\bar{L}}\rparen^{-1}\tilde{\TDOAMatrix}_{\mbf{L}}\,\mathds{1}+\mathds{1}\tilde{\TDOAMatrix}_{\mbf{L}}\lparen\mbf{D_\beta}+\mbf{\bar{L}}\rparen^{-1}
\end{multline}

This completes the proof.
\end{IEEEproof}

\ifthenelse{\equal{\useAlgNames}{true}}
{
From now on, we will refer to this algorithm as \MatrixCompletionAlg.
}

It is noteworthy to comment that the matrix
$(\mbf{D_\beta}+\mbf{\bar{L}})$ contains important information about the
recoverability of missing data: if it is full-rank, then the solution of
\eqref{eq:matrix_completion} is unique and if
$(\mbf{D_\beta}+\mbf{\bar{L}})$ is rank-deficient, missing data is not recoverable uniquely without any further assumption.

Furthermore, in the absence of missing data,
$n(\mbf{D_\beta}+\nobreak\mbf{\bar{L}})^{-1}=\nobreak\mbf{I}$, hence the matrix
completion solution in \eqref{eq:closed-form-completion} becomes the
solution of the denoising problem stated in \eqref{eq:closed-form}.






\section{Robust TDOA Denoising with Missing Data}
\label{sec:robust-missing}

In this section we aim to combine the results of
sections~\ref{sec:robust-denoising} and~\ref{sec:missing-data},
addressing the more general case in which both outliers and missing data
are considered. Therefore, the problem is a combination of
\eqref{eq:robust-denoising} and \eqref{eq:matrix_completion} defined as:
 \begin{equation}
  \label{eq:robust-missing}
  \begin{aligned}
    & \underset{\TDOAMatrix,\mbf{S}}{\operatorname{minimize}} & &
    \left\|\mbf{L}\hp\lparen\tilde{\TDOAMatrix}-\TDOAMatrix-\mbf{S}\rparen\right\|_F^2 \\
    & \operatorname{subject\;to} & & \TDOAMatrix\in\TDOASETn\\
    & & & \|\mbf{S}\|_0 < 2k\\
    & & & \mbf{S} = \mbf{L}\hp\mbf{S}.
  \end{aligned}
\end{equation}

In the same way as in section~\ref{sec:robust-denoising},
(\ref{eq:robust-missing}) can be solved by alternatively solving the
following two subproblems until convergence:

\begin{subequations}
  \label{eq:robust-missing-godec}
    \begin{align}
      \TDOAMatrix_t &= \underset{\TDOAMatrix\in\TDOASETn}{\operatorname{arg\,min}}\quad
      \left\|\mbf{L}\hp\lparen\tilde{\TDOAMatrix}-\TDOAMatrix-\mbf{S}_{t-1}\rparen\right\|_F^2\label{eq:robust-missing-godec1}\\
      \mbf{S}_t &= \underset{\|\mbf{S}\|_0<2k}{\operatorname{arg\,min}}\quad
      \left\|\mbf{L}\hp\lparen\tilde{\TDOAMatrix}-\TDOAMatrix_t\rparen-\mbf{S}\right\|_F^2.\label{eq:robust-missing-godec2}
    \end{align}
\end{subequations}
The subproblem \eqref{eq:robust-missing-godec1} is equivalent to the
missing data problem solved in section~\ref{sec:missing-data} but
considering
$\tilde{\TDOAMatrix}_{\mbf{L}}=\nobreak({\mbf{L}\hp\tilde{\TDOAMatrix}}-\nobreak\mbf{S}_{t-1})$. Therefore,
according to theorem~\ref{th:closed-form-completion}, it has a closed
form solution:
  \begin{multline}
  \label{eq:robust-missing-sol1}
  \TDOAMatrix_t^{\ast}=(\mbf{D_\beta}+\mbf{\bar{L}})^{-1}(\mbf{L}\hp\tilde{\TDOAMatrix}-\mbf{S}_{t-1})\,\mathds{1}+\\+\mathds{1}(\mbf{L}\hp\tilde{\TDOAMatrix}-\mbf{S}_{t-1})(\mbf{D_\beta}+\mbf{\bar{L}})^{-1}.
\end{multline}
Since \eqref{eq:robust-missing-godec2} is of the same form as the second
subproblem in \eqref{eq:modified_godec}, it can also be solved by
entry-wise hard thresholding of
$\mbf{L}\hp(\tilde{\TDOAMatrix}-\TDOAMatrix_t)$.

The pseudocode shown in Alg.~\ref{alg:rodenmis} summarizes the proposed
algorithm for the general case.

\begin{algorithm}[H]
  \caption{Robust denoising with missing data.}\label{alg:rodenmis}
  \textbf{Require:} $\tilde{\TDOAMatrix}$, $\mbf{L}$, $k$, $\epsilon$ 

  \textbf{Ensure:} $\TDOAMatrix\in\TDOASETn$, $\|\mbf{S}\|_0<2k$, 

  $\enspace$\footnotesize{1:}$\enspace$\normalsize  $\mbf{D_\beta}=\mbf{I}\hp\mbf{L}\mbf{L}^\top$

  $\enspace$\footnotesize{2:}$\enspace$\normalsize  $\mbf{Q}=(\mbf{D_\beta}+\mbf{\bar{L}})^{-1}$ \label{lin:pre-cal}

  $\enspace$\footnotesize{3:}$\enspace$\normalsize  $\TDOAMatrix_0=\tilde{\TDOAMatrix}$ ; $\mbf{S}_0=0$ ; $t=0$

  $\enspace$\footnotesize{4:}$\enspace$\normalsize  \textbf{while} $\|\tilde{\TDOAMatrix}-\TDOAMatrix_t-\mbf{S}_t\|_F^2/\|\tilde{\TDOAMatrix}\|_F^2<\epsilon$ \textbf{do}

  $\enspace$\footnotesize{5:}$\enspace$\normalsize $\quad$  $t=t+1$

  $\enspace$\footnotesize{6:}$\enspace$\normalsize $\quad$  $\TDOAMatrix_t=\mbf{Q}(\mbf{L}\hp\tilde{\TDOAMatrix}-\mbf{S}_{t-1})\,\mathds{1}+\mathds{1}(\mbf{L}\hp\tilde{\TDOAMatrix}-\mbf{S}_{t-1})\mbf{Q}$

  $\enspace$\footnotesize{7:}$\enspace$\normalsize $\quad$  $\mbf{S}_t=\mathcal{P}_{2k}(\tilde{\TDOAMatrix}-\TDOAMatrix_t)$ 

  $\enspace$\footnotesize{8:}$\enspace$\normalsize \textbf{end while}

  $\enspace$\footnotesize{9:}$\enspace$\normalsize \textbf{return} $\TDOAMatrix_t$, $\mbf{S}_t$
\end{algorithm}

Note that in line~\ref{lin:pre-cal} the matrix
$\mbf{Q}=(\mbf{D_\beta}+\mbf{\bar{L}})^{-1}$ can be precalculated in
order to get an efficient implementation of the algorithm.

\ifthenelse{\equal{\useAlgNames}{true}}
{
From now on, we will refer to this algorithm as \RobustDenoisingMatrixCompletionAlg.
}

\section{Experiments with Synthetic Data}
\label{sec:synthetic-data}

In this section computer simulations will be used to compare the
proposed algorithms with some of the alternatives existing in the state
of the art. 


For evaluating the \RobustDenoisingAlg{} and
\RobustDenoisingMatrixCompletionAlg{} algorithms, two different metrics
will be used:

\begin{itemize}
\item The Signal-to-Noise-Ratio SNR [dB] of the non-redundant set
  referenced to the first sensor
  ($10\log(\sum_{i=1}^n{\|\Delta\tau_{i1}\|^2/\sum_{i=1}^n{\|\Delta{\tau}_{i1}^\ast-\Delta\tau_{i1}\|^2}})$). This
  is an application independent metric (where $\Delta{\tau}_{i1}^\ast$
  is the estimation of $\Delta\tau_{i1}$), that will allow assessing
  the proposal improvements in the TDOA measurements \emph{per se}.
\item The localization error, measured as the average distance between
  the source ground truth position and the position estimated using
  any given localization algorithm based on TDOA estimations (such as
  \cite{chan1994simple} in our case). This is an application dependent
  metric, that will allow assessing the actual benefits of the
  proposal in an example of a real task. Note that our proposal is not
  restricted to localization and can be used in other applications
  that could benefit from denoised TDOAs (such as self-calibration or
  beamforming).
\end{itemize}

\subsection{Experimental setup}
\label{sec:experimental-setup-syn}

For all the synthetic data experiments, a set of 10 sensors (which
implies 45 different sensor pairs) and 1 source were randomly
located. Therefore, 45 different TDOA measurements were generated per
experiment, and independent Gaussian noise was added to them, using the
same variance for all the measurements.

The sensor locations were uniformly distributed in a cube of 1 meter
side, and the source positions were uniformly distributed in a 2 meter
side cube. The propagation speed of the signal was set to 343.313
m/s. In all the experiments where it's required, $\epsilon$ is set to
$10^{-10}$. To increase the statistical significance of the results, they
  are provided as averages of 20 independent runs.

\subsection{Evaluation of Robust TDOA Denoising}
\label{sec:synth-denoising}

In this first experiment, we evaluated the performance of the
\RobustDenoisingAlg{} algorithm proposed in
section~\ref{sec:robust-denoising}, imposing that some TDOA values were
outliers. To simulate this, we randomly chose some measurements (between
0 and 10) and replaced them with a zero-mean Gaussian distributed noise,
with a standard deviation of 0.1 ms. It is worth mentioning that the
outlier values calculated that way are not related at all to the real
TDOAs, thus being \emph{true} outliers.  The parameter $k$ of the
proposed algorithm, which fixes the maximum number of identifiable
outliers, was set to 8.

\subsubsection{SNR Improvements Evaluation}
\label{sec:snr-impr-eval}

\begin{figure}[!t]
  \centering
  \includegraphics[width=0.49\textwidth]{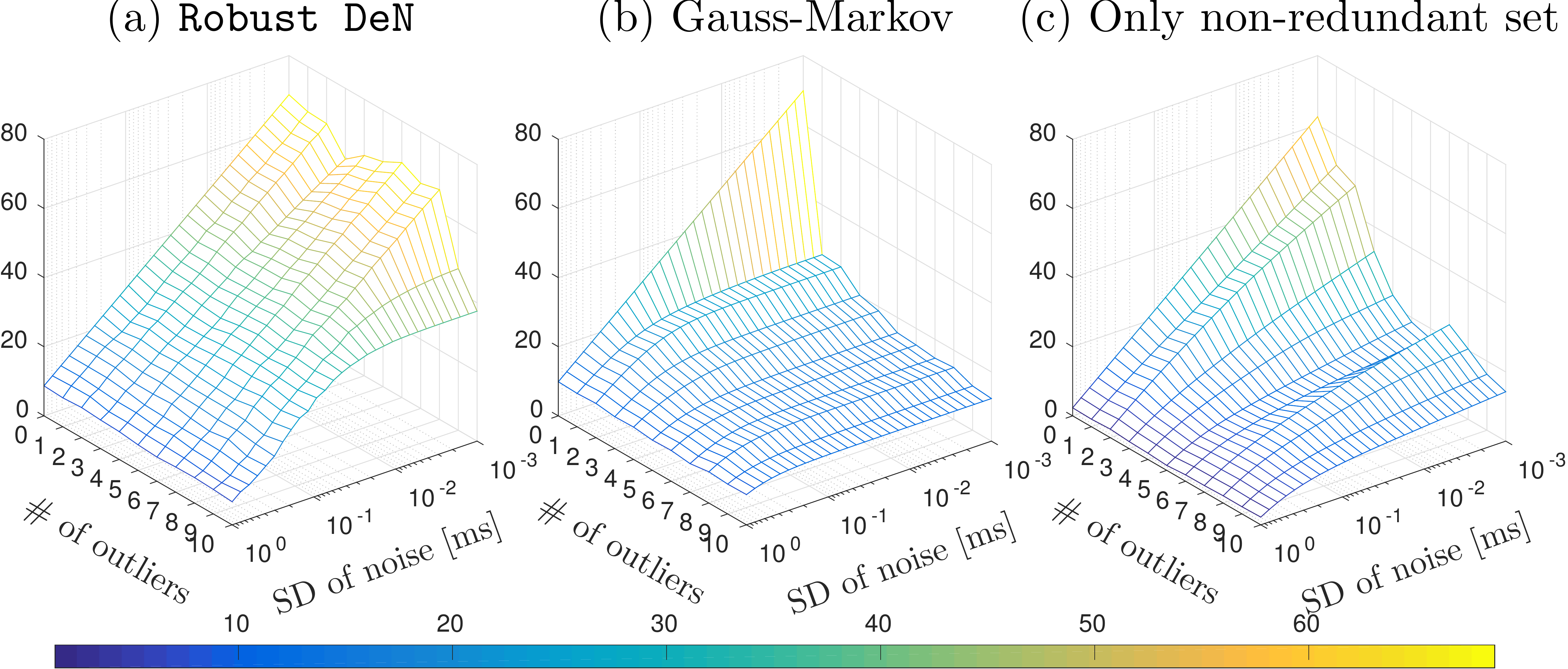}
  \caption{Robust denoising in synthetic data: SNR in dB, higher is better.}
  \label{fig:MapSNR_synthetic_dn_10mics}
\end{figure}

Fig.~\ref{fig:MapSNR_synthetic_dn_10mics}a shows the SNR values for the
\RobustDenoisingAlg{} algorithm when modifying the noise
standard deviation and the number of outliers, compared with that
obtained by the Gauss-Markov estimator
(Fig.~\ref{fig:MapSNR_synthetic_dn_10mics}b), and also when only the non
redundant set is used, i.e. not using the redundancy of TDOA
measurements (Fig.~\ref{fig:MapSNR_synthetic_dn_10mics}c).

As predicted in section~\ref{sec:denoising}, when no outliers are
present, the performance of the \RobustDenoisingAlg{} algorithm is the
same as Gauss-Markov (see row 0 in
Figs.~\ref{fig:MapSNR_synthetic_dn_10mics}a
and~\ref{fig:MapSNR_synthetic_dn_10mics}b), hence it reaches the
Cramer-Rao Bound \cite{so2008closed}, while being much better than using
no redundancy.  Nevertheless, the proposed algorithm clearly outperforms
the other two approaches when outliers are present in the measurements
(rows 1 through 10 in the graphics of
Fig.~\ref{fig:MapSNR_synthetic_dn_10mics}).

\subsubsection{Source Localization Improvements Evaluation}
\label{sec:source-local-impr}

\begin{figure}[!t]
  \centering
  \includegraphics[width=0.48\textwidth]{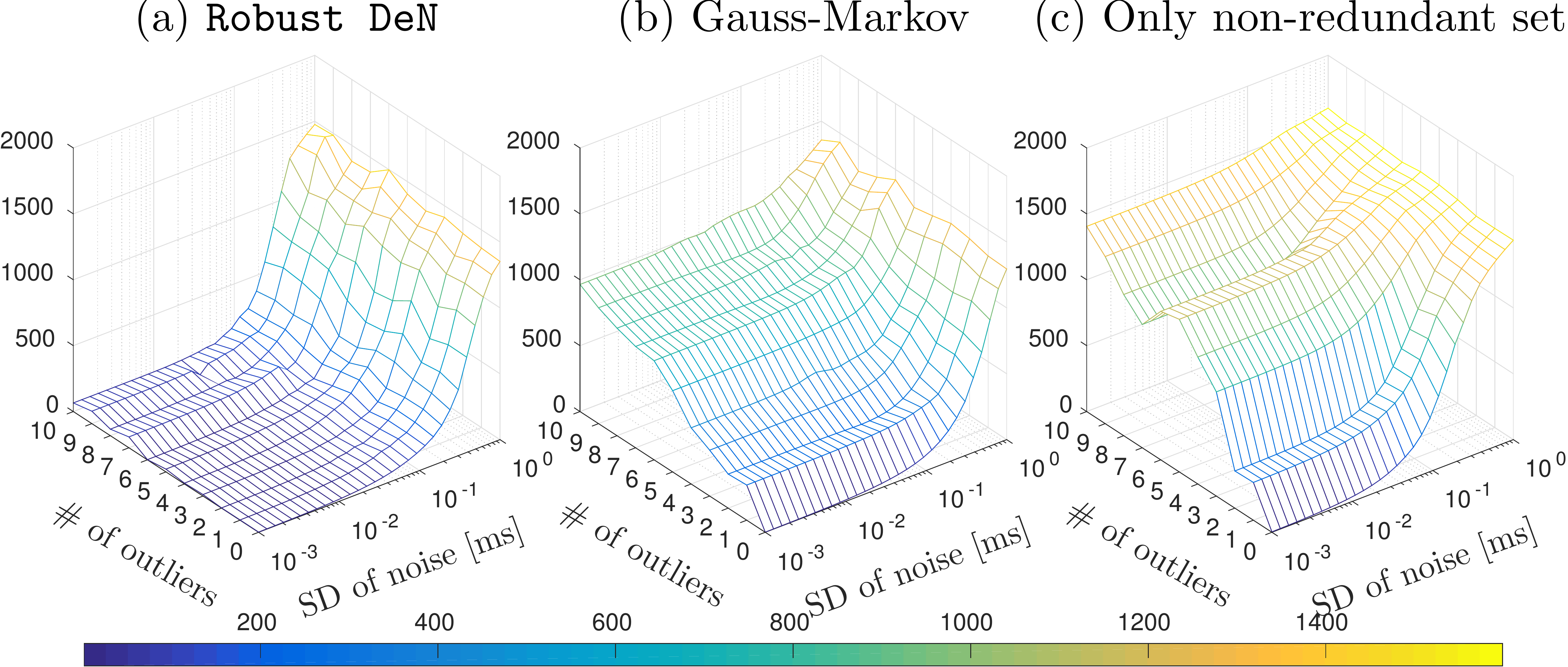}
  \caption{Robust denoising in synthetic data: Localization error in mm (using
    \cite{chan1994simple}), lower is better.}
  \label{fig:MapCHAN_synthetic_dn_10mics}
\end{figure}

The optimized non redundant set provided by the algorithms applied in
Section~\ref{sec:snr-impr-eval} were used in a localization algorithm
using \cite{chan1994simple}. The average localization errors (in mm) are
shown in Fig.~\ref{fig:MapCHAN_synthetic_dn_10mics}. Again, the
\RobustDenoisingAlg{} algorithm performs as Gauss-Markov when there are no
outliers, but is clearly superior when outliers are present.

It is also worth mentioning that the behaviour of the robust denoising
keeps the improvements at roughly the same level for increasing number
of outliers present, thus validating the ability of the algorithm to
pinpoint and eliminate their presence.

\subsection{Evaluation of Missing Data Recovery}
\label{sec:synth-completion}

In this second experiment, we evaluated the capability of the
\MatrixCompletionAlg{} algorithm proposed in
section~\ref{sec:missing-data} to recover missing values. For our
purposes, the missing TDOA measurements were also chosen randomly but,
in contrast to the previous experiment, the matrix positions of the
missing measurements were known.

\begin{figure}[!t]
  \centering 
   \includegraphics[width=0.48\textwidth]{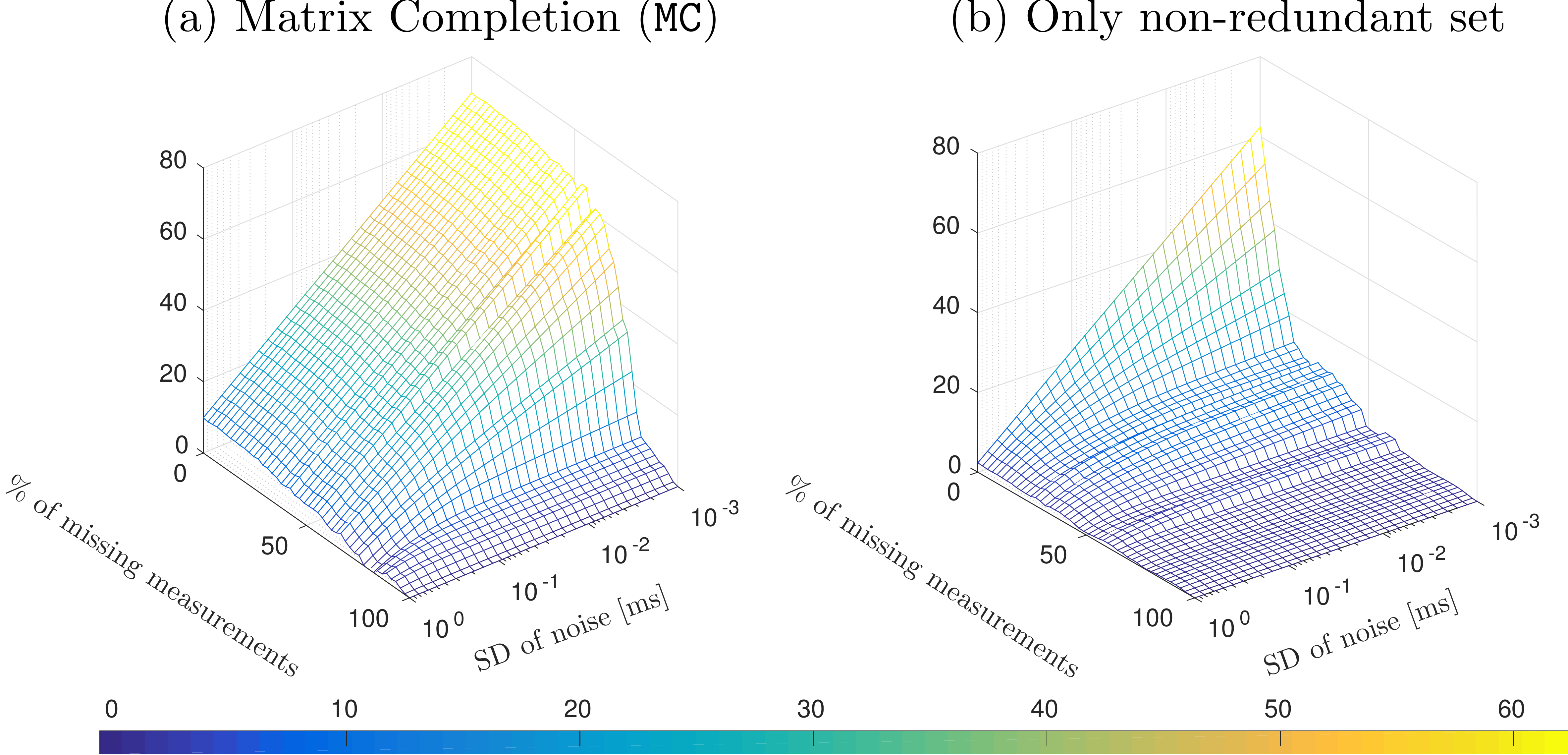}
  \caption{Missing data recovery in synthetic data: SNR in dB, higher
    is better.}
  \label{fig:MapSNR_synthetic_comp_10mics}
\end{figure}
\begin{figure}[!t]
  \centering 
   \includegraphics[width=0.48\textwidth]{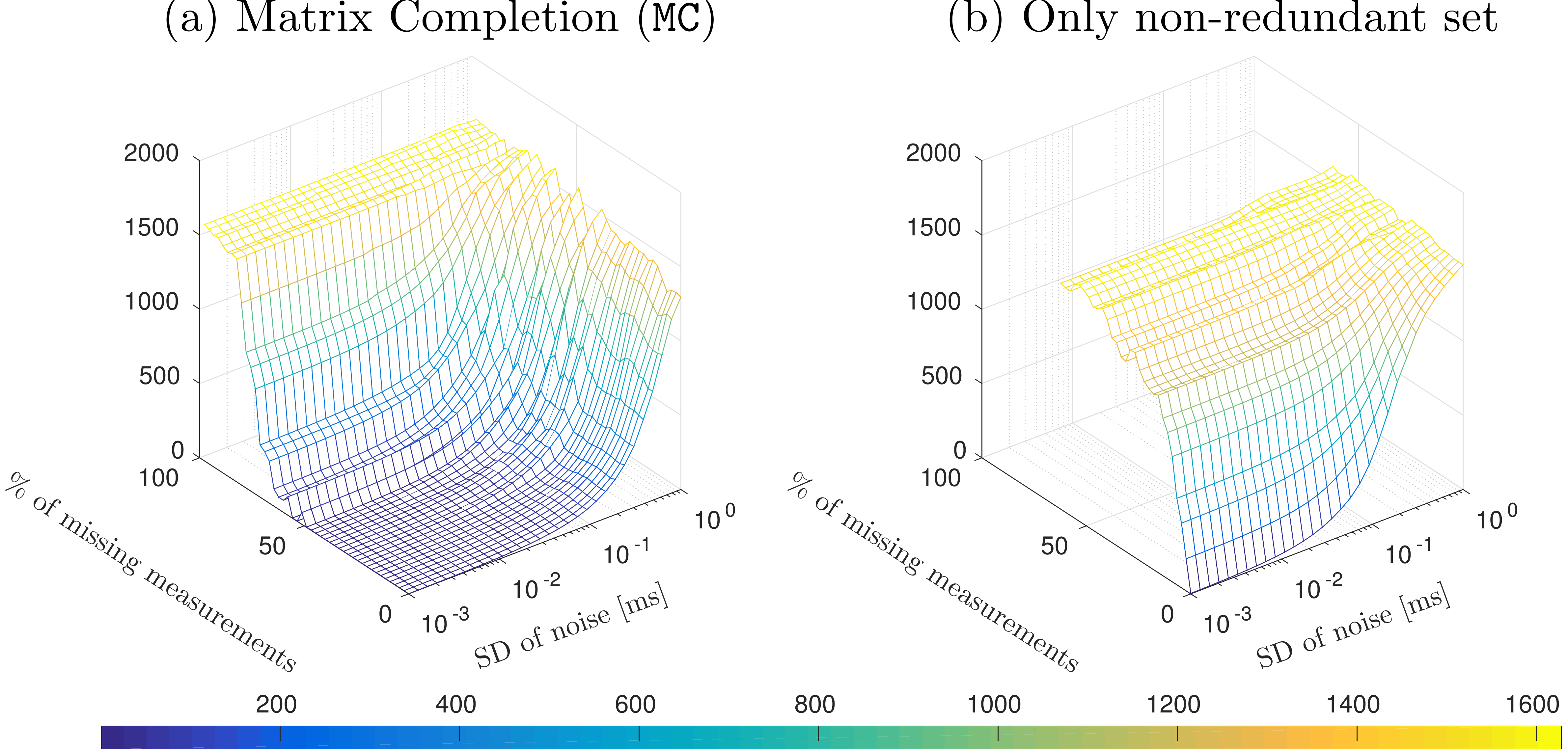}
  \caption{Missing data recovery in synthetic data: Localization error
    in mm (using \cite{chan1994simple}), lower is better.}
  \label{fig:MapCHAN_synthetic_comp_10mics}
\end{figure}

Fig.~\ref{fig:MapSNR_synthetic_comp_10mics} and
Fig.~\ref{fig:MapCHAN_synthetic_comp_10mics} show, respectively, the SNR
values, and the localization error for the \MatrixCompletionAlg{}
algorithm, when modifying the noise standard deviation and the
percentage of missing TDOA values in the TDOA matrix, as compared with
using only the non-redundant set (missing values were set to zero).

From the figures, it can be clearly seen that the proposed algorithm can
take advantage of the knowledge about which measurements were missing,
achieving even better results than when the positions of the outliers
were unknown. For example, removing 50\% of the full TDOA set of measurements 
implies 23 missing values of 45 measurements , much more than the maximum of 10 outliers
evaluated in Fig.~\ref{fig:MapSNR_synthetic_dn_10mics}, while keeping
good performance.

\subsection{Evaluation of Robust TDOA Denoising with Missing Data}

In this third experiment, we evaluated the capability of the
\RobustDenoisingMatrixCompletionAlg{} algorithm
proposed in section~\ref{sec:robust-missing} to face both outliers and
recover missing values.

To provide a wide range of evaluation scenarios, we defined: \emph{i)}
Two conditions related to noise, namely \emph{low} and \emph{high}. The
former corresponds to a standard deviation of $10^{-3}$ ms., and the
latter to $0.2$ ms. \emph{ii)} Two conditions related to the presence of
outliers, imposing the existence of 2 or 6 outliers. \emph{iii)} A
variable number of missing TDOA measurements, defined as a percentage of
missing TDOA values in the TDOA matrix.

In all cases, the number of outliers is fixed among the full
TDOA set and then, some measurements are discarded, i.e. some of
the discarded measurements may be outliers. Note that this is consistent
with the real case, where it is not possible to anticipate where the
outliers are.

\begin{figure*}[!t]
  \centering 
  \begin{subfigure}[t]{0.18\textwidth}
    \includegraphics[width=\textwidth]{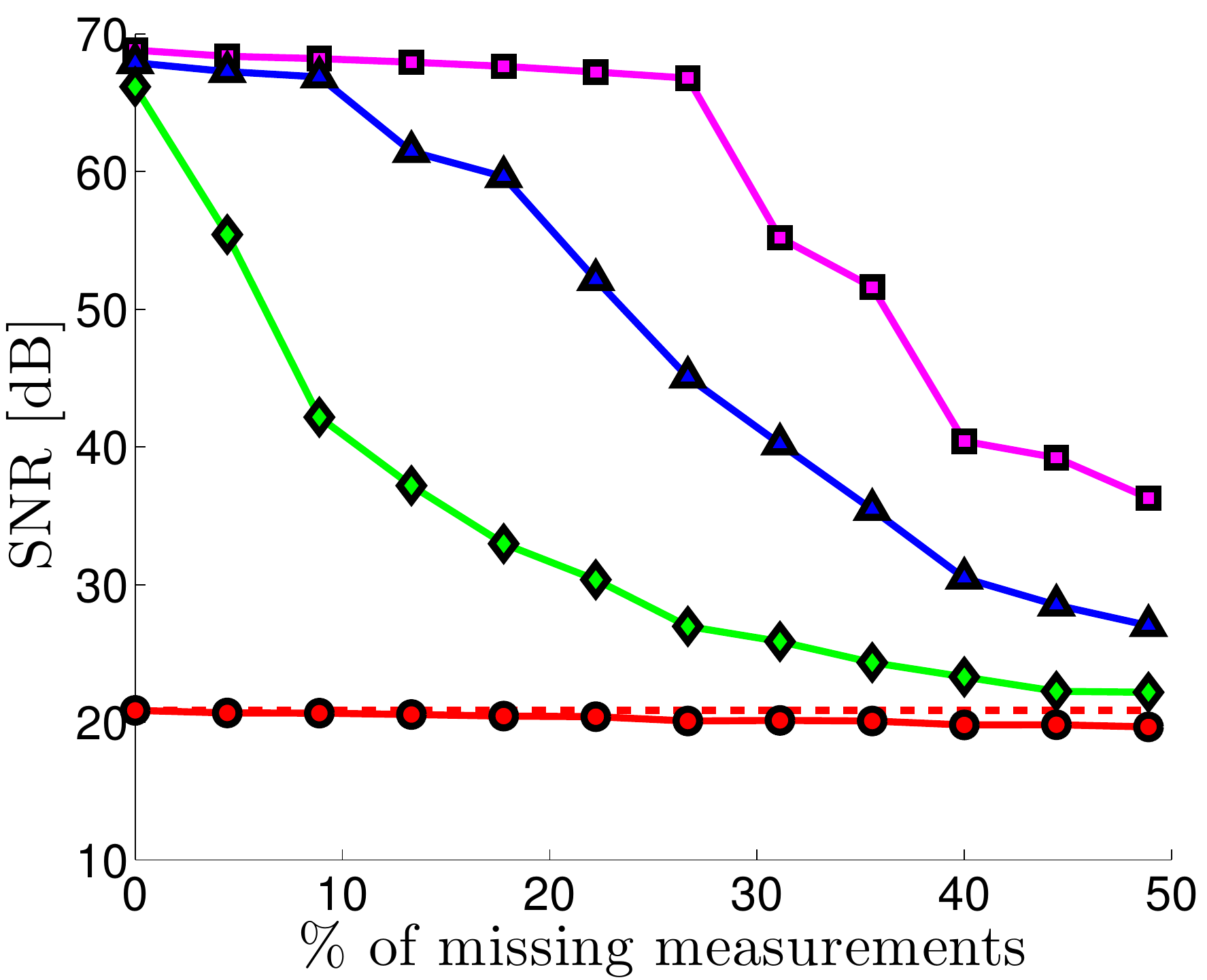}
    \caption{SNR with low noise and 2 outliers}
    \label{fig:SNR_synth_low_few}
  \end{subfigure}
\;
  \begin{subfigure}[t]{0.18\textwidth}
    \includegraphics[width=\textwidth]{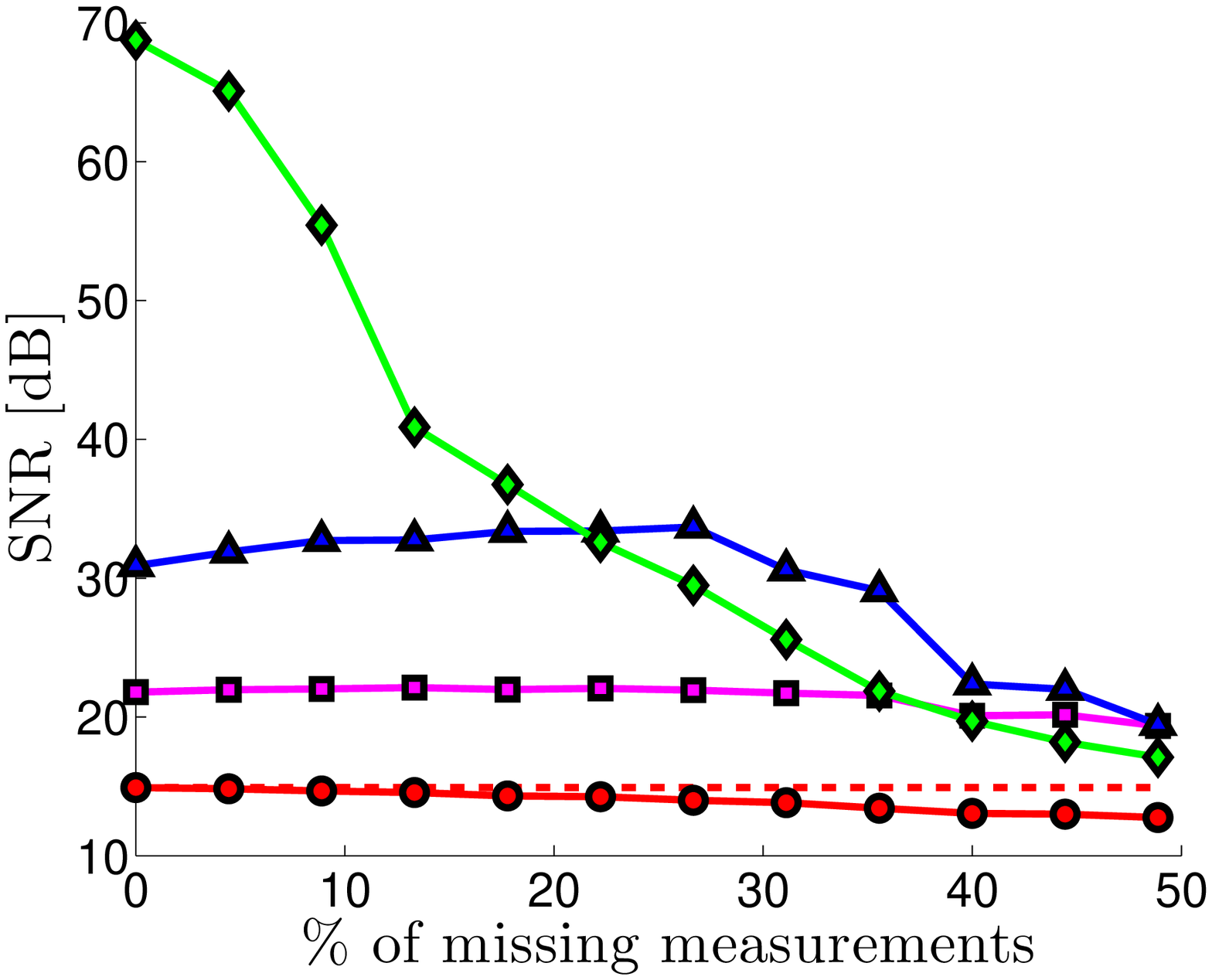}
    \caption{SNR with low noise and 6 outliers}
    \label{fig:SNR_synth_low_many}
  \end{subfigure}
\;
  \begin{subfigure}[t]{0.18\textwidth}
    \includegraphics[width=\textwidth]{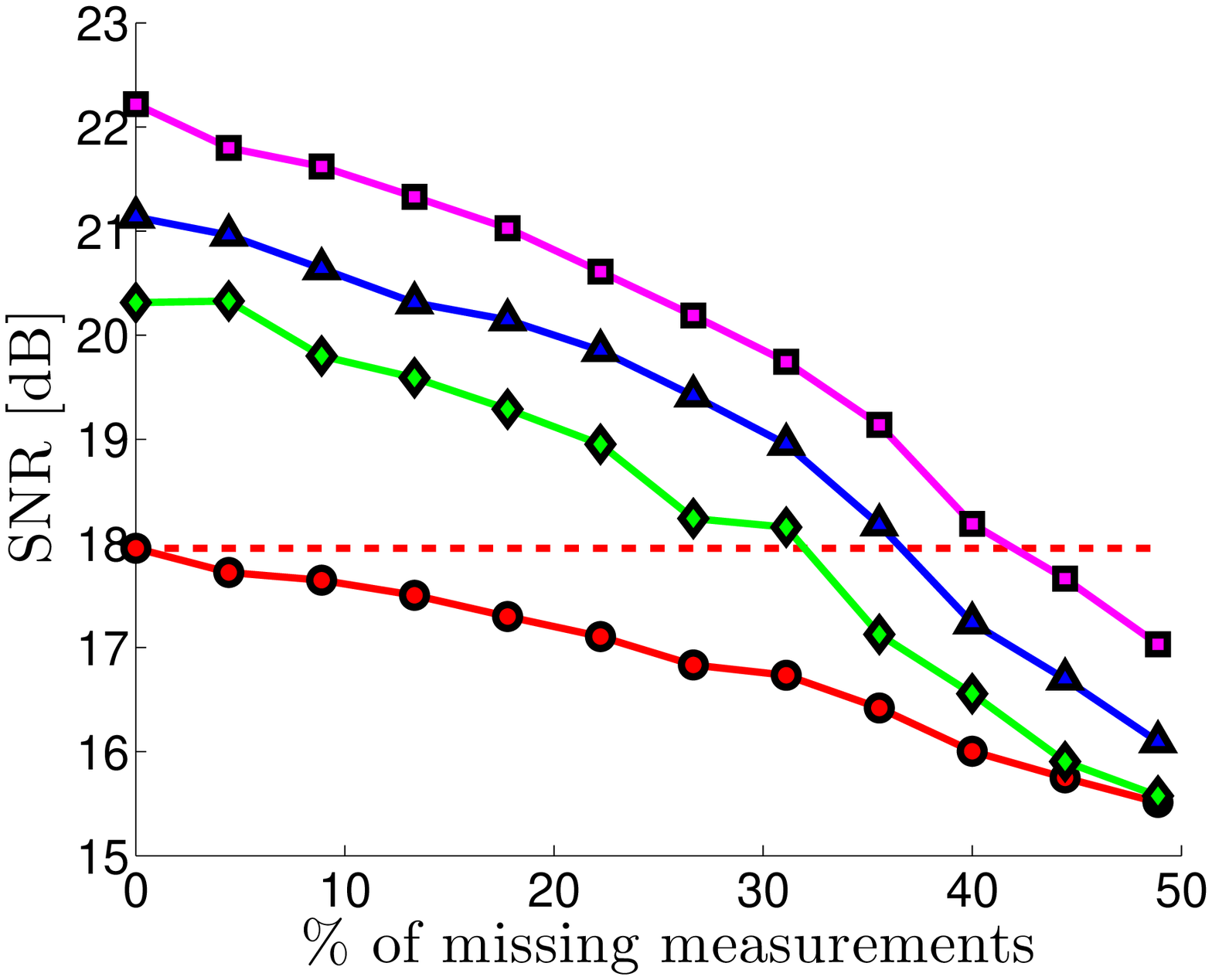}
    \caption{SNR with high noise and 2 outliers}
    \label{fig:SNR_synth_high_few}
  \end{subfigure}
\;
  \begin{subfigure}[t]{0.18\textwidth}
    \includegraphics[width=\textwidth]{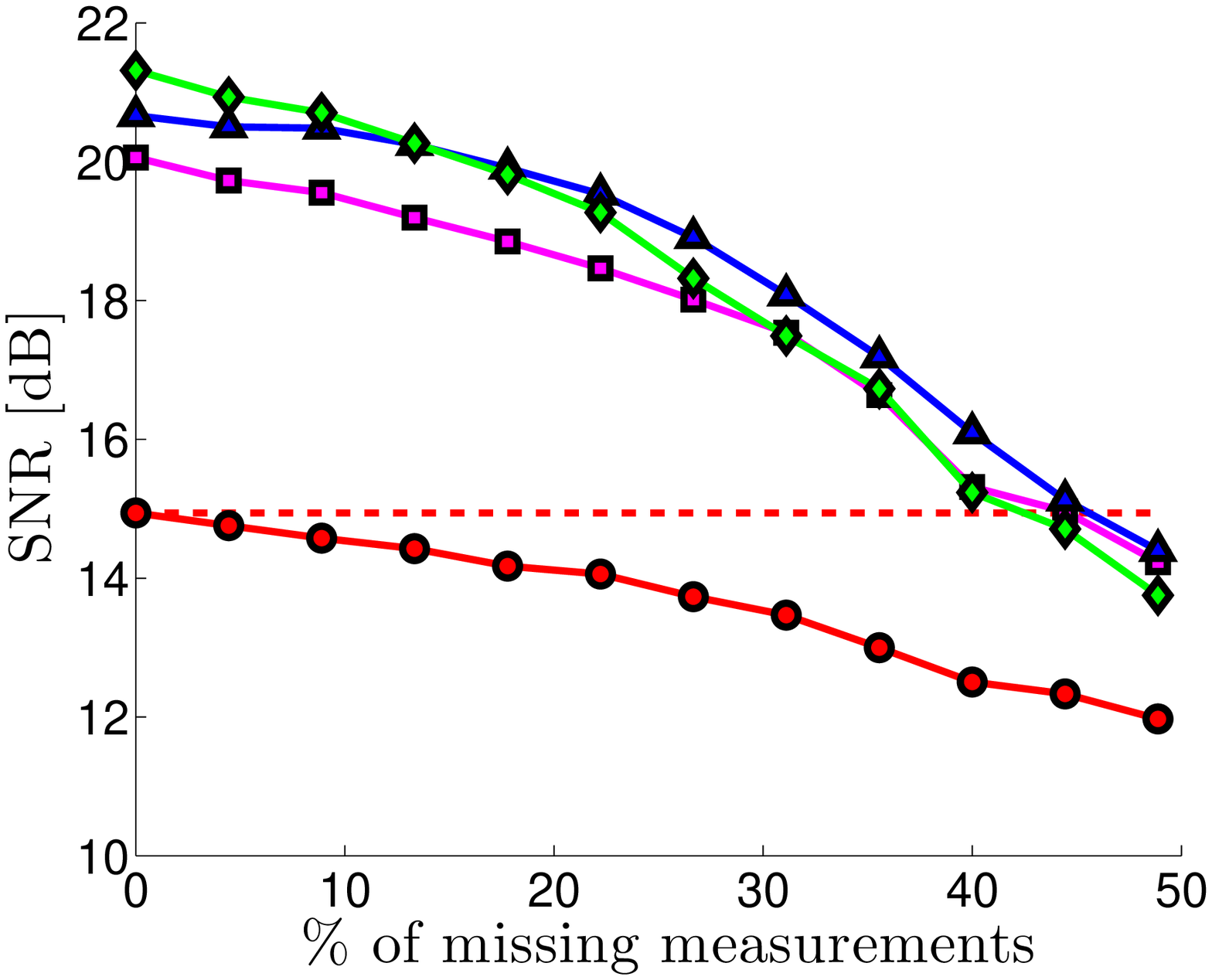}
    \caption{SNR with high noise and 6 outliers}
    \label{fig:SNR_synth_high_many}
  \end{subfigure}
  \begin{subfigure}[t]{0.18\textwidth}
    \includegraphics[width=\textwidth]{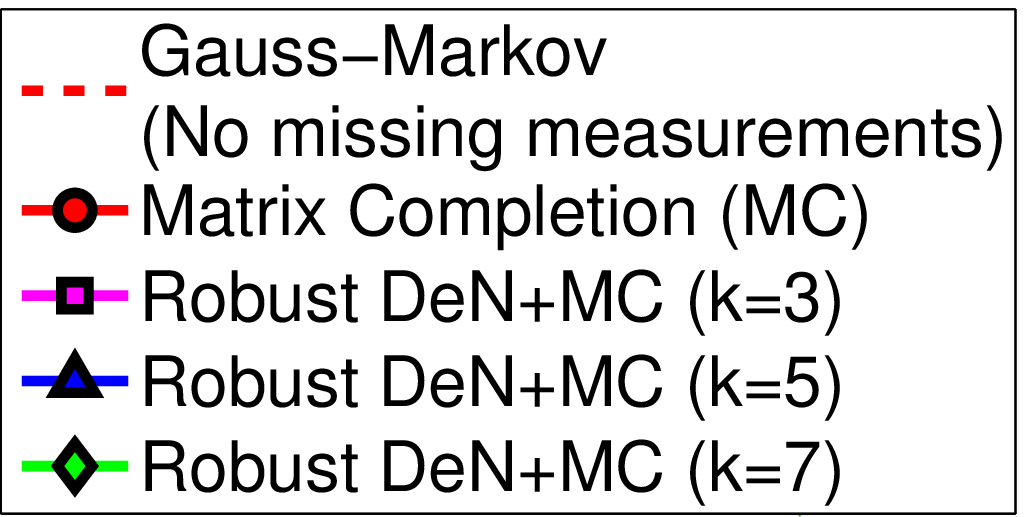}
  \end{subfigure}

  \caption{Algorithm evaluation in synthetic data: SNR in dB.}
  \label{fig:SNR_comparative_synth}
\end{figure*}

\begin{figure*}[!t]
  \centering 
  \begin{subfigure}[t]{0.18\textwidth}
    \includegraphics[width=\textwidth]{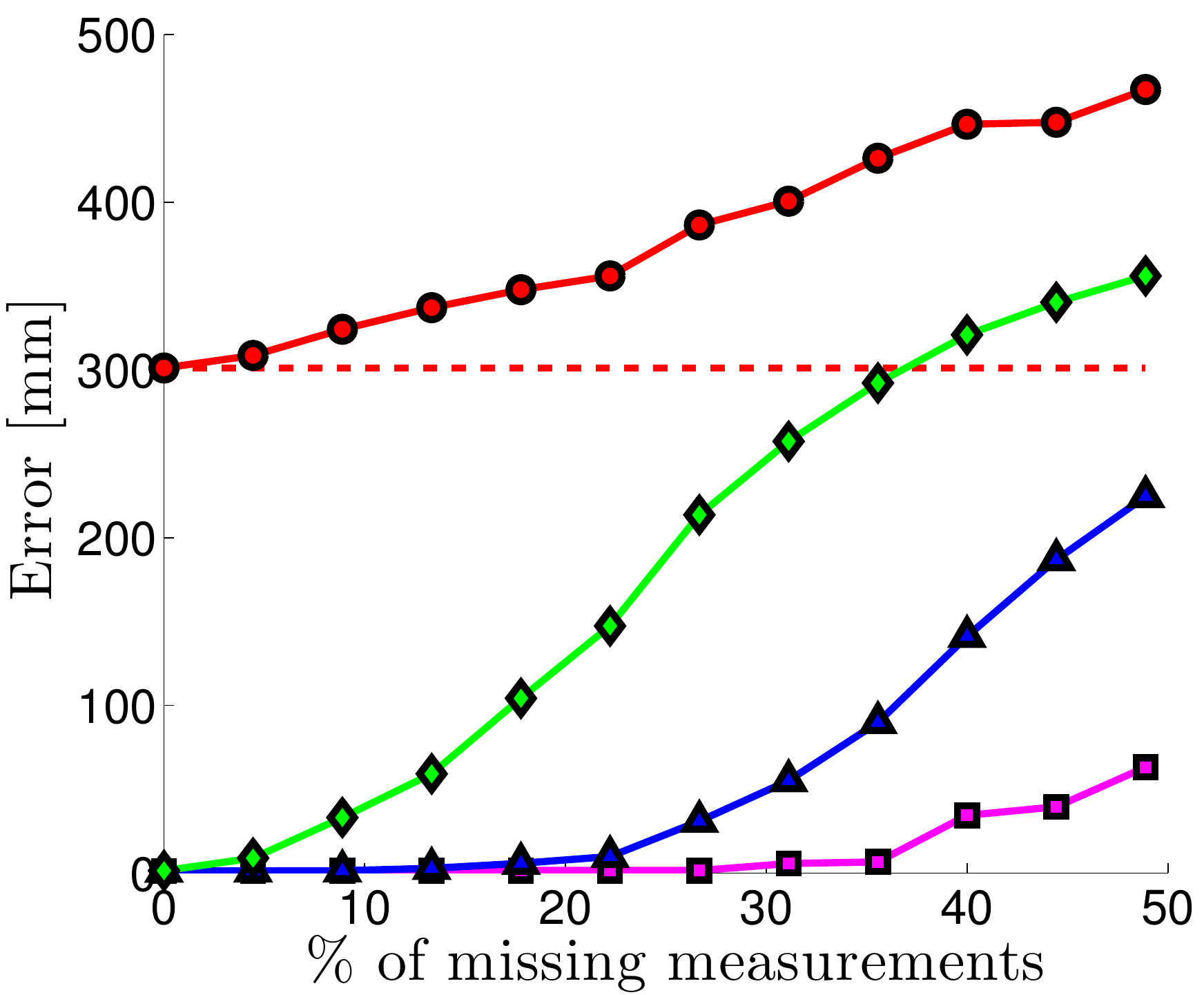}
    \caption{Localization error with low noise and 2 outliers}
    \label{fig:Error_synth_low_few}
  \end{subfigure}
\;
  \begin{subfigure}[t]{0.18\textwidth}
    \includegraphics[width=\textwidth]{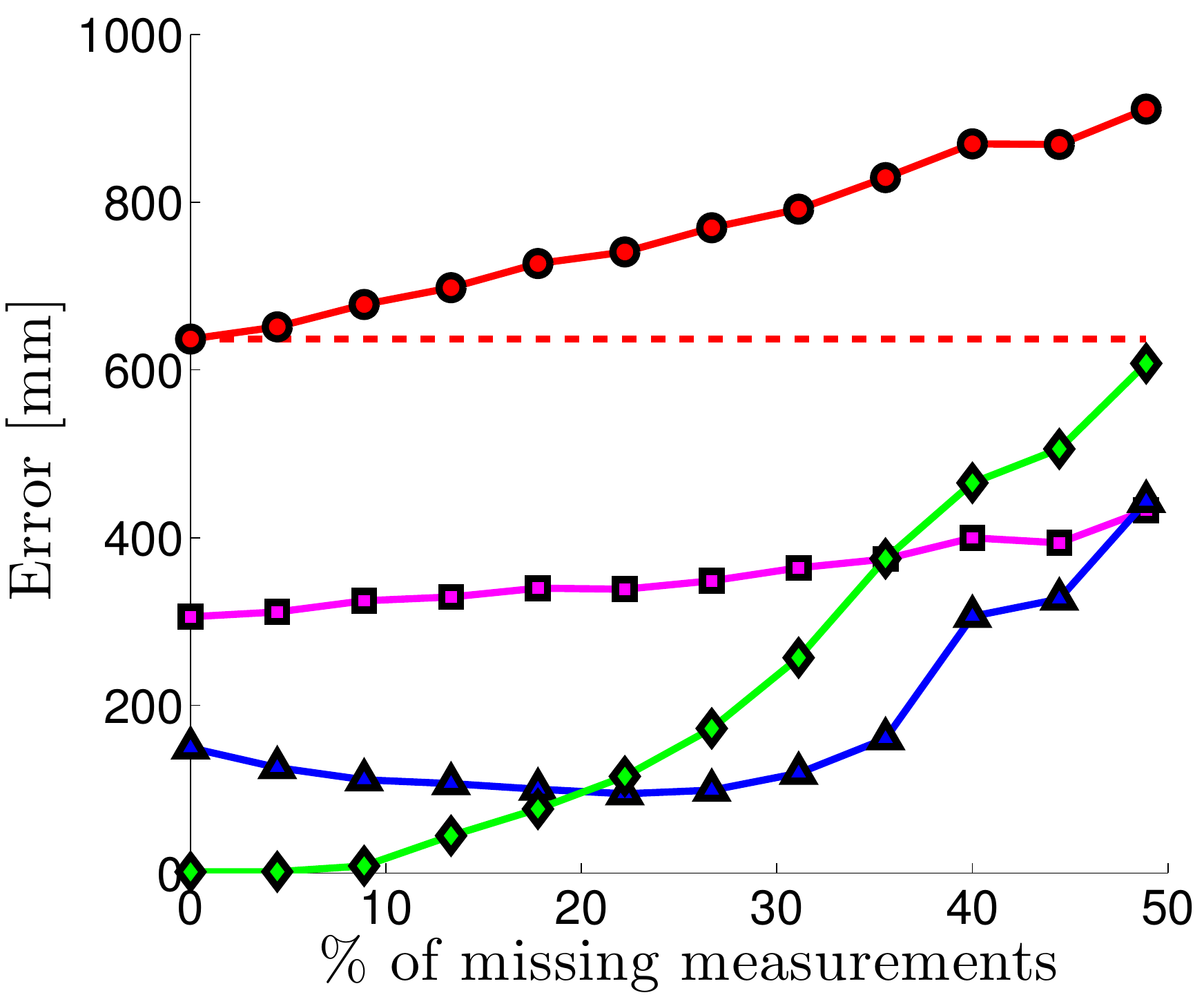}
    \caption{Localization error with low noise and 6 outliers}
    \label{fig:Error_synth_low_many}
  \end{subfigure}
\;
  \begin{subfigure}[t]{0.18\textwidth}
    \includegraphics[width=\textwidth]{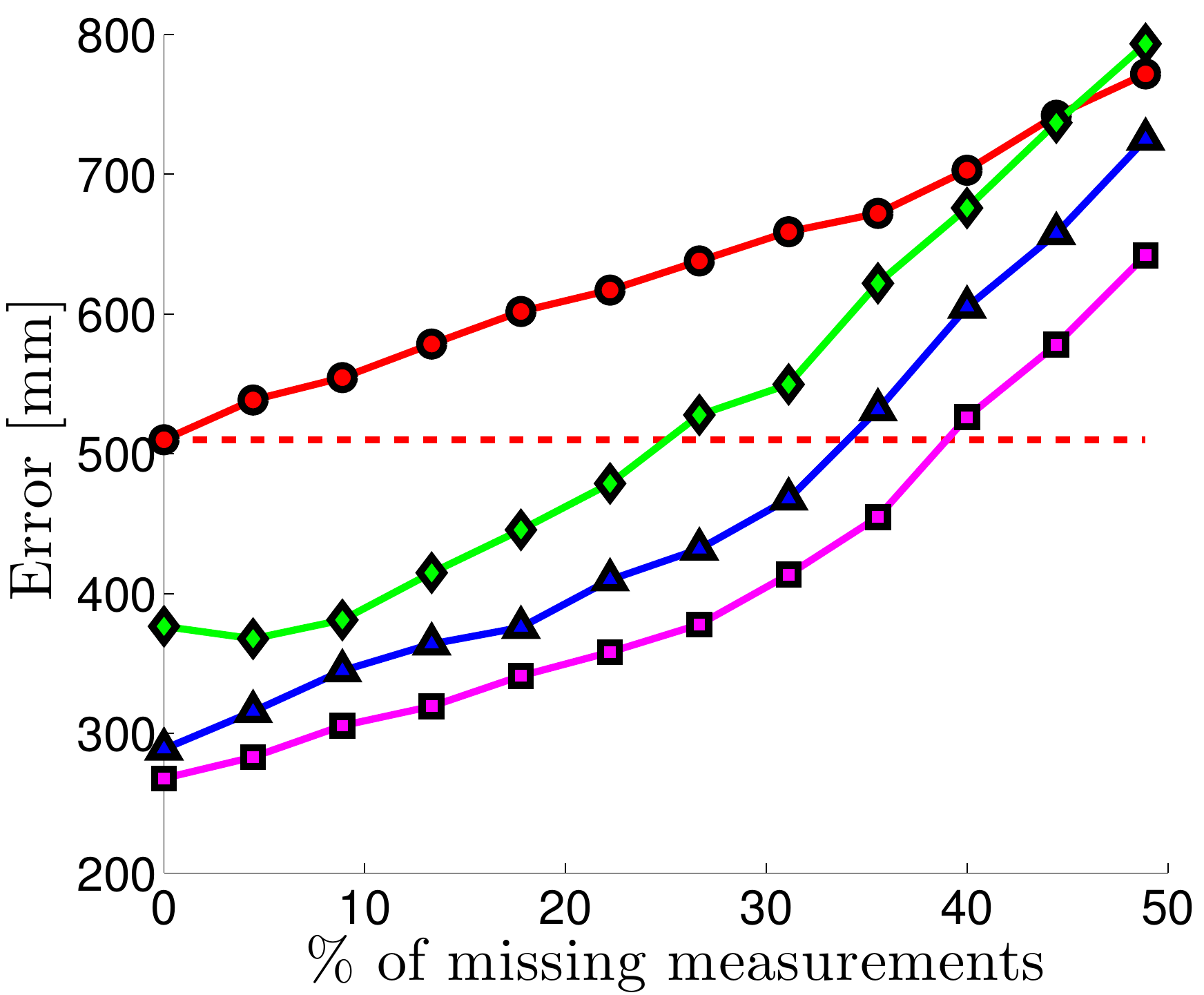}
    \caption{Localization error with high noise and 2 outliers}
    \label{fig:Error_synth_high_few}
  \end{subfigure}
\;
  \begin{subfigure}[t]{0.18\textwidth}
    \includegraphics[width=\textwidth]{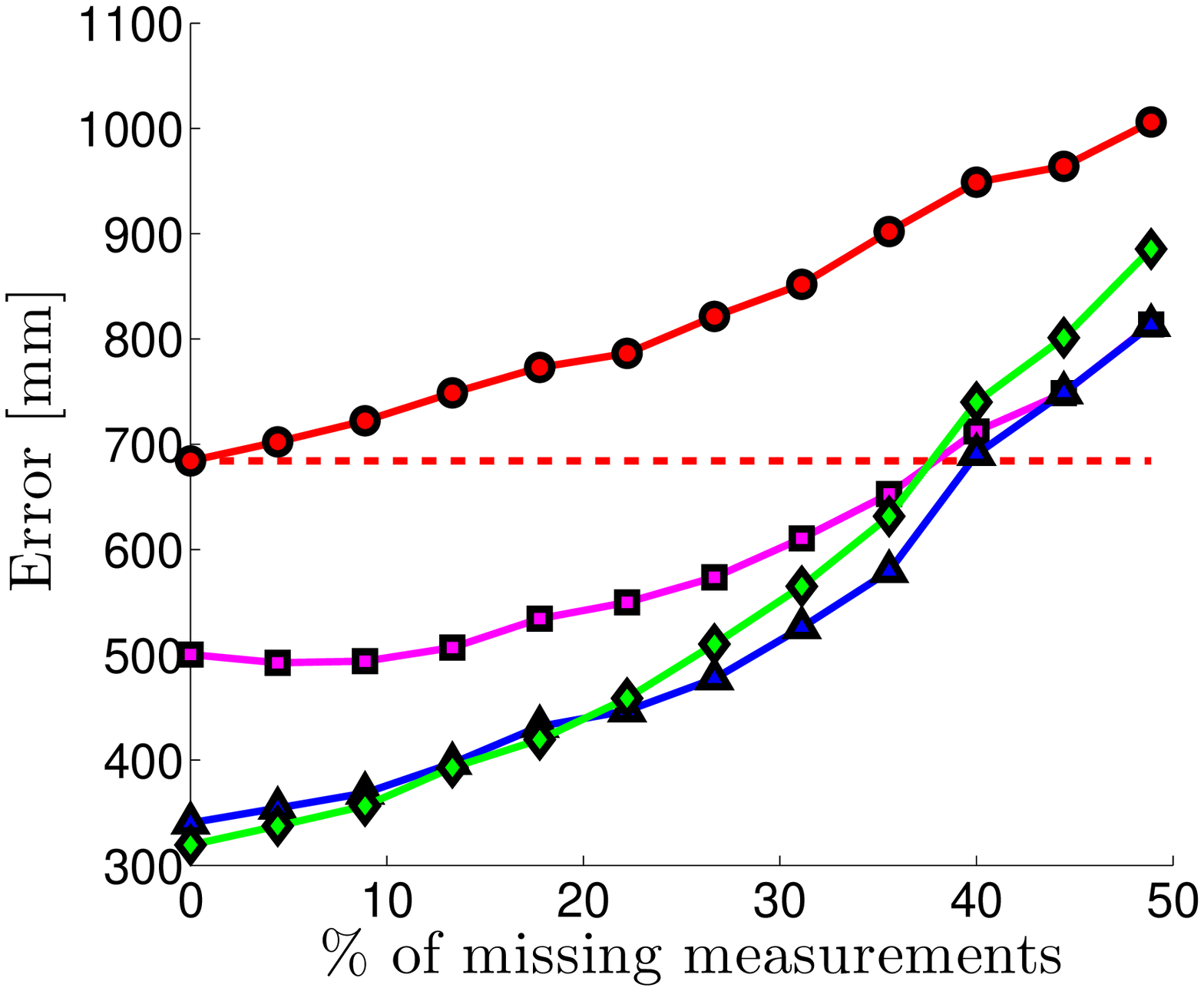}
    \caption{Localization error with high noise and 6 outliers}
    \label{fig:Error_synth_high_many}
  \end{subfigure}
\;
  \begin{subfigure}[t]{0.18\textwidth}
    \includegraphics[width=\textwidth]{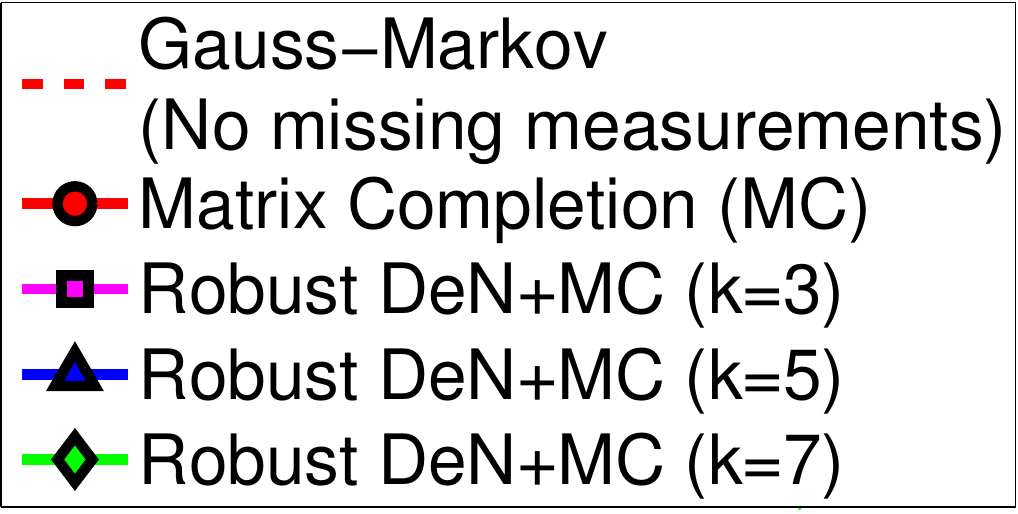}
  \end{subfigure}

  \caption{Algorithm evaluation in synthetic data: Localization error in
    mm (\cite{chan1994simple} is used for
    source position estimation from the optimized TDOA values).}
  \label{fig:Error_comparative_synth}
\end{figure*}

Fig.~\ref{fig:SNR_comparative_synth}
and Fig.~\ref{fig:Error_comparative_synth} show, respectively, the SNR
values, and the localization error for different algorithms, and for
different evaluation scenarios.

As it can be seen in Fig.~\ref{fig:SNR_synth_low_few},
\ref{fig:SNR_synth_high_few}, \ref{fig:Error_synth_low_few}, and
\ref{fig:Error_synth_high_few}, when there are a low number of outliers
(2 in this case), the best results are obtained for lower $k$ values.
However, when the number of outliers increase,
(Fig.~\ref{fig:SNR_synth_low_many}, \ref{fig:SNR_synth_high_many},
\ref{fig:Error_synth_low_many} and \ref{fig:Error_synth_high_many}, low
$k$ values perform worse. So, we can conclude that $k$ must be a number
as low as possible, but higher than the number of actual outliers.

Nevertheless, it is worth to observe the behaviour of
Fig.~\ref{fig:SNR_synth_low_many}, \ref{fig:SNR_synth_high_many},
\ref{fig:Error_synth_low_many} and \ref{fig:Error_synth_high_many} (with
more outliers) when the percentage of missing data increases. It can be
clearly seen that the lines corresponding to different values of $k$ are
crossing among them. This seems to indicate that as the missing data
percentage increases, the number of outliers that we are able to detect
decreases.

Anyway, the results obtained by the
\RobustDenoisingMatrixCompletionAlg{} algorithm outperforms the
Gauss-Markov estimator, asymptotically approaching it when the noise is
very high. Note also that for high values of noise, the noise and the
outliers are practically indistinguishable.

\section{Experiments with Real Data}
\label{sec:real-data}


The aim of this section is to evaluate whether the improvements obtained
in section~\ref{sec:synthetic-data} using synthetic data are actually
found in real environments. To do this, the proposed algorithms have
been evaluated using audio recordings from the AV16.3
database~\cite{lathoud2005av16}, an audio-visual corpus recorded in
the \emph{Smart Meeting Room} of the IDIAP Research Institute, in
Switzerland. The same metrics and localization algorithm of the
previous section has been employed.

Additionally, the proposed \RobustDenoisingAlg{} algorithm is compared
with other recent state-of-the-art methods in
section~\ref{sec:comparison}. In that case, we have employed the same
framework used in~\cite{alameda2014geometric}, wherein many methods were
already compared. In order to make the comparison as fair as possible,
in this part the localization algorithm will be the same as
in~\cite{alameda2014geometric}.

\subsection{Experimental Setup}
\label{sec:validation}

\begin{figure}
  \centering
  \begin{subfigure}[t]{0.24\textwidth}
    \includegraphics[width=0.8\textwidth]{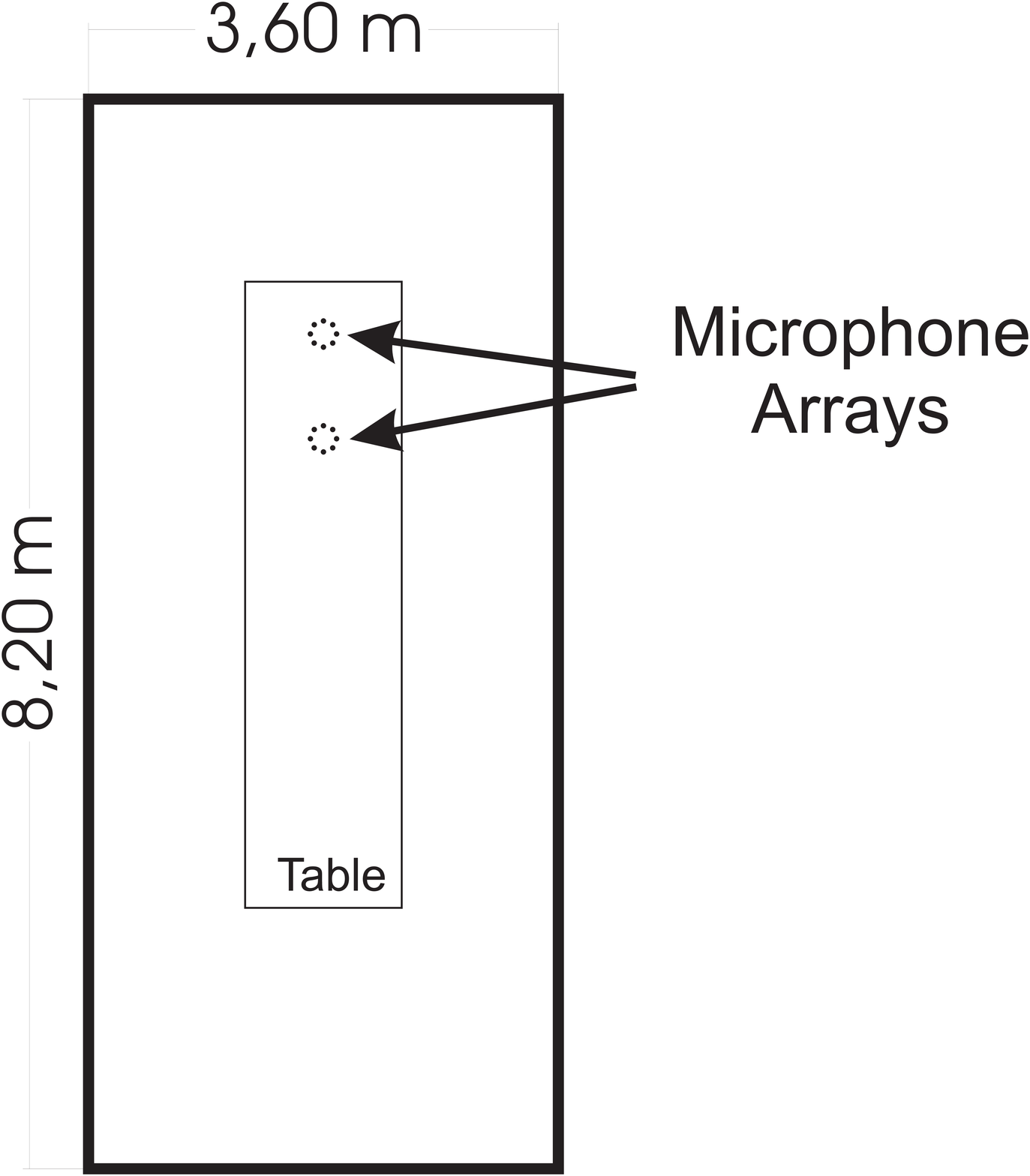}
    \caption{IDIAP room layout showing the centered table, and the
      microphones arranged in two circular arrays.}
    \label{fig:RoomLayout}
  \end{subfigure}%
\qquad
  \begin{subfigure}[t]{0.17\textwidth}
    \includegraphics[width=\textwidth]{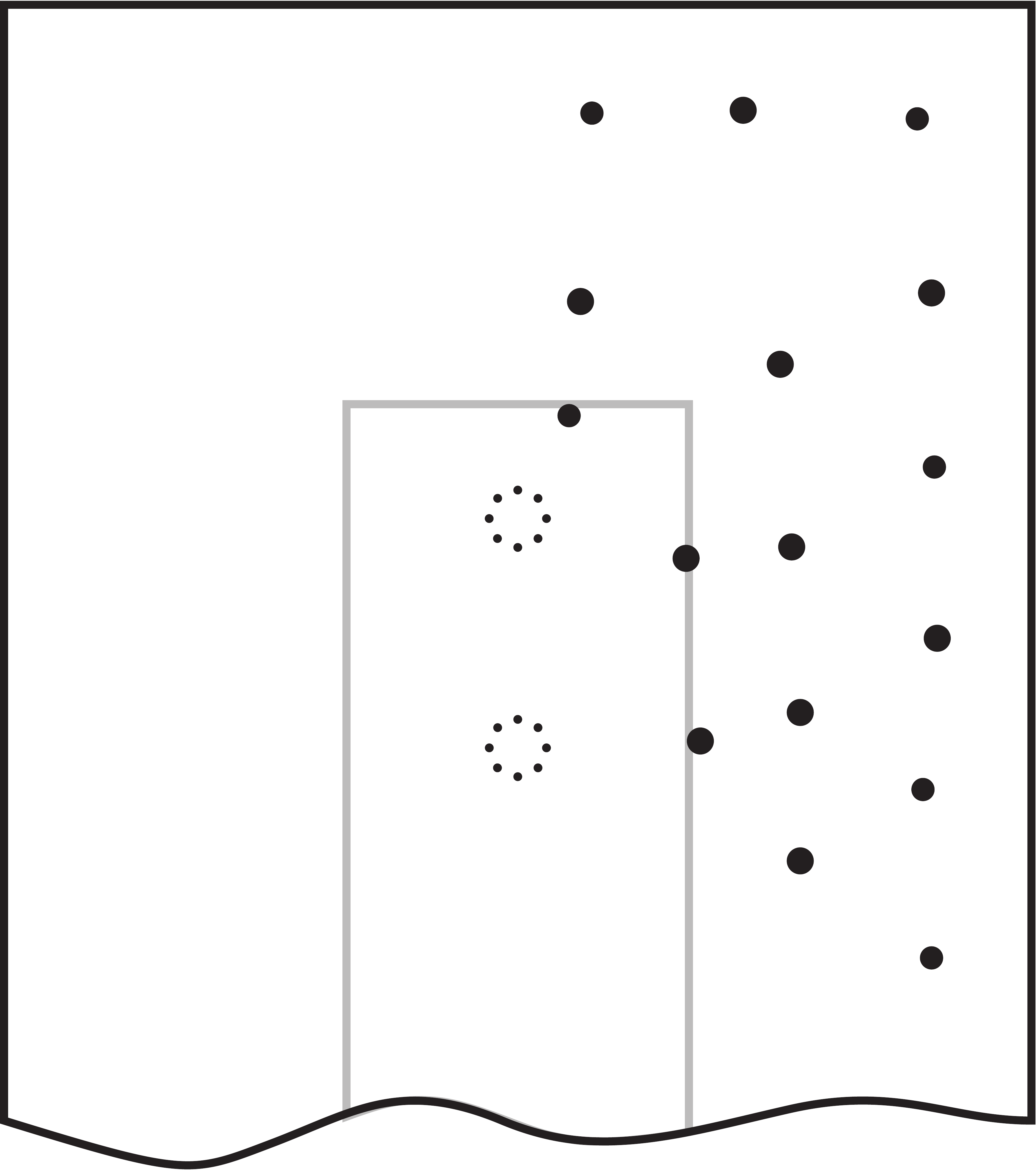}
    \caption{Positions evaluated in the real data experiments. Only the
      relevant room section is shown.}
    \label{fig:real_positions_short}
  \end{subfigure}
  \caption{IDIAP Smart Meeting Room: experimental details.}
  \label{fig:LIdiapRoom}
\end{figure}

The IDIAP Meeting Room (shown in Fig.~\ref{fig:LIdiapRoom}) is a
$8.2m \times \nobreak 3.6m \times 2.4m$ rectangular space containing a
centrally located $4.8m \times 1.2m$ rectangular table, on top of
which two circular microphone arrays of $10 cm$ radius are located,
each of them composed by 8 microphones. The centers of the two arrays
are separated by $80 cm$ and the origin of coordinates is located in
the middle point between the two arrays. A detailed description of the
meeting room can be found in~\cite{moore2002}.

The audio recordings are synchronously sampled at $16~KHz$, and the
complete database along with the corresponding annotation files
containing the recordings ground truth (3D coordinates of the speaker's
mouth) is fully accessible on-line at~\cite{av163}. It is composed by
several sequences from which we are using sequence 01, with a single
male speaker generating digit strings in 16 positions (which can be seen
as small circles in Fig.~\ref{fig:real_positions_short}), distributed
along the room. The sequence duration accounts for 208 seconds in total,
with 823 ground truth frames.

The TDOA measurements $\Delta\tilde{\tau}_{ij}$, from which the measured
TDOA matrix $\mbf{\tilde{M}}$ is built, where estimated using the
highest peak of the GCC-PHAT function\cite{knapp1976GCC}.

As in a real scenario outliers are common but difficult to anticipate or
enforce, the sweep over noise levels and the number of outliers that we
performed with synthetic data are not feasible. Therefore, in our
experiments with real data, we will only provide the SNR values and
localization errors obtained after using each algorithm.



\subsection{Evaluation of Robust TDOA Denoising}
\label{sec:source-local-impr2}

In this experiment, \textit{all} the microphone pairs have been
considered, hence 120 TDOA values have been computed for each frame.

In table~\ref{tab:results-real-data} we show an example of the results
for the \RobustDenoisingAlg{} algorithm with $k=10$. As it also happened with synthetic data, in this case the proposed algorithm outperforms the Gauss-Markov estimator, yielding great improvements in both SNR and localization precision. Fig.~\ref{fig:comparative_real} shows that the selection of $k$ is not very critical in this dataset as improvements over Gauss-Markov are obtained for a wide range of $k$. 

\begin{table}[t!]
  \renewcommand{\arraystretch}{1.3}
  \caption{Robust denoising performance in real data}
  \label{tab:results-real-data}
  \centering
  \begin{tabular}{l c c}
    \hline
    &SNR (dB)&Average Localization error (mm)\\
    \hline
    \RobustDenoisingAlg{} & 27.46  &354\\
    Gauss-Markov& 23.19 &515\\
    Only non-redundant set&17.83 &858
  \end{tabular}.
\end{table}

These results are the baseline for the experiments with missing data
described in the next subsection.

\subsection{Evaluation of Robust TDOA Denoising with Missing Data}
\label{sec:deno-with-miss}

\begin{figure}[!t]
  \centering 
  \begin{subfigure}[t]{0.24\textwidth}
    \includegraphics[width=\textwidth]{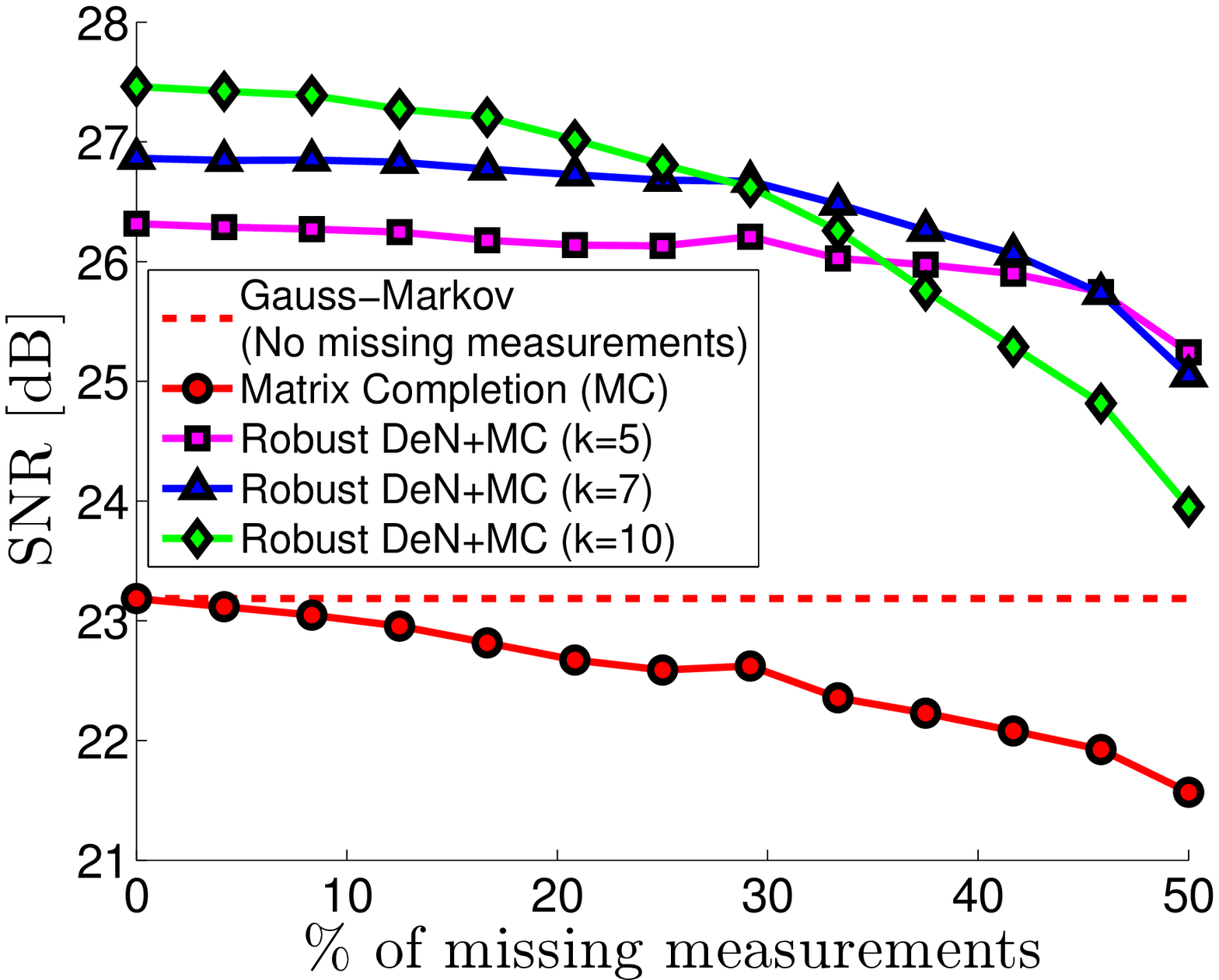}
    \caption{SNR in dB}
    \label{fig:Error_real}
  \end{subfigure}
  \begin{subfigure}[t]{0.24\textwidth}
    \includegraphics[width=\textwidth]{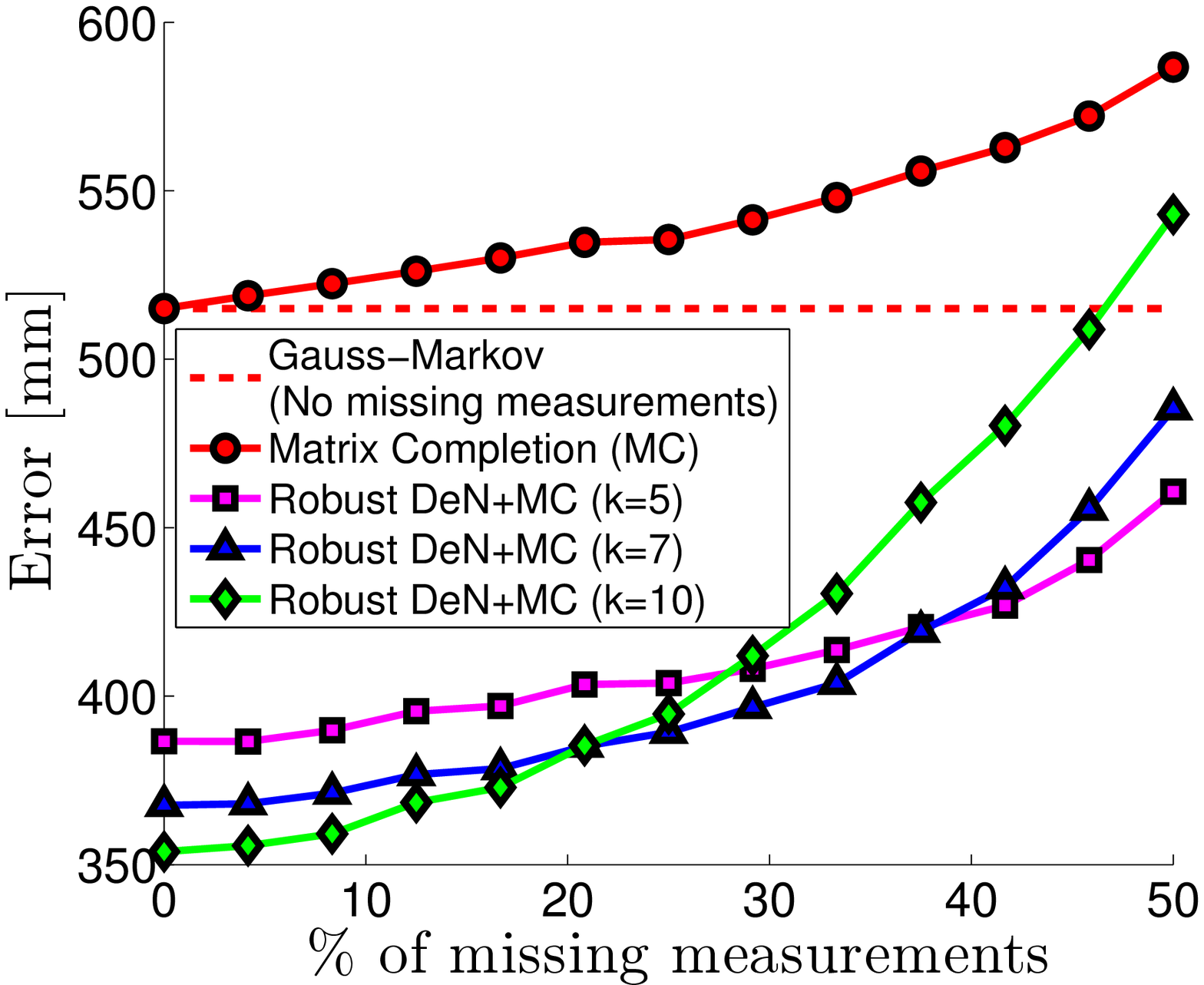}
    \caption{Localization error in mm (using  \cite{chan1994simple})}
    \label{fig:SNR_real}
  \end{subfigure}

  \caption{Results for real data with missing TDOA measurements.}
  \label{fig:comparative_real}
\end{figure}

In the second experiment with real data, we randomly remove a set of
TDOA measurements. Fig.~\ref{fig:comparative_real} shows the obtained
results.  The dotted lines correspond to the performance (SNR and
localization error) achieved by the Gauss-Markov estimator when there
are no missing measurements. The solid lines with circular marks are the
results obtained by the \MatrixCompletionAlg{} algorithm described in
section~\ref{sec:missing-data}.

On the other hand, the solid lines with squared/triangular/diamond
marks correspond with the results of the
\RobustDenoisingMatrixCompletionAlg{} algorithm presented in
section~\ref{sec:robust-missing}. The different colors/shapes indicate
different values of the hyperparameter $k$.

Fig.~\ref{fig:comparative_real} highlights the relevance of the proposed
robust-denoising algorithm (\RobustDenoisingMatrixCompletionAlg{}) in
real-scenarios, with important improvements over its non robust version
(\MatrixCompletionAlg{}): higher than 4dB absolute improvement in terms
of SNR, and around 30\% relative improvement (15 cm absolute) in terms
of localization precision.

We again observe that as the percentage of missing data increases, the
lines corresponding to different values of $k$ are crossing among
them. This behaviour is very similar to that found in the synthetic
experiments above (refer to Fig.~\ref{fig:SNR_comparative_synth}
and Fig.~\ref{fig:Error_comparative_synth}) for a high number of outliers,
what suggests that this is the case in the real experiment, also
serving as a validation for our simulation conclusions.

It is also noteworthy that, in order to get the best result, the
maximum number of outliers $k$ should be decreased when the number of
missing measurements increases.

\subsection{Comparison with Other Methods in a Localization Task}
\label{sec:comparison}

In this section we made use of the code and real data of the gtde MATLAB
toolbox~\cite{alameda2014gtde,alameda2014geometric}. In this toolbox,
real recordings were performed under noisy conditions in a 4x4x4 (m)
room, in which an array of 4 microphones forming a tetrahedron of 20 cm
side has been placed.  Sound waves coming from a loudspeaker placed at
189 different locations 1.7 m away from the array were recorded at 48
KHz.

Table~\ref{tab:results-real-data2} shows the results of the different
algorithms on the localization task. The first 5 columns refer to the
results of the \RobustDenoisingAlg{} algorithm for 5 different values of
the parameter $k$. The rest of the columns show the results of a
selection of the algorithms implemented and evaluated
in~\cite{alameda2014geometric} (the nomenclature has been kept and the
results are essentially the same). 

From~\cite{alameda2014geometric}, we selected three multilateration
methods (generically referred to as \emph{x-mult}) which are
implementations of the algorithm described
in~\cite{Brandstein97practical}. These methods require to be
initialized with the distance $r$ to the source\footnote{\emph{n-mult},
  \emph{t-mult} and \emph{f-mult}, use $r$ values of $0.9m$, $1.7m$ and
  $2.5m$ respectively, following~\cite{alameda2014geometric}.}. They
were selected because they also make use of the redundancy between all
the correlations, using the same input as the \RobustDenoisingAlg{}
algorithm. We also compare our proposal with the branch \& bound
(\emph{b\&b}) method proposed by the authors
of~\cite{alameda2014geometric}, as this is the one with the best
performance in that work. The interested reader may refer
  to~\cite{alameda2014geometric} for further details about the used
  methods, and the results of some other methods as well.

The \RobustDenoisingAlg{} algorithm is the only one of the evaluated
algorithms that does not require information about the array geometry to
perform denoising of the TDOA estimations. The localization algorithm
(i.e converting TDOA to angles), implemented in~\cite{alameda2014gtde},
was used after all the methods in order to make the comparison as fair
as possible.

The first row in table~\ref{tab:results-real-data2} shows the percentage
of localization measurements with an angular error lower than 30º, the
second row shows their mean angular error, and the third row their
standard deviation.


\begin{table*}[t!]
  \renewcommand{\arraystretch}{1.3}
  \caption{Real data performance comparison between
    \RobustDenoisingAlg{} and selected algorithms
    in~\cite{alameda2014geometric} (marked with $^\ast$) ($T_{60}\approx0.5s$)}
  \label{tab:results-real-data2}
  \centering
  \begin{tabular}{lccccccccc}
    \hline
    &\multicolumn{5}{c}{\RobustDenoisingAlg{}}&\multirow{2}{*}{\emph{n-mult}$^\ast$}&\multirow{2}{*}{\emph{t-mult}$^\ast$}&\multirow{2}{*}{\emph{f-mult}$^\ast$}&\multirow{2}{*}{\emph{b\&b}$^\ast$}\\

    &$k=0$  &$k=1$  &$k=2$  &$k=3  $&$k=4$  & & & & \\
    \hline
   \% Measurements with angular error $< 30^\circ$ 
    &23.78\%&24.12\%&25.20\%&25.71\%&23.52\%&13.99\%&17.38\%&17.91\%&27.60\%\\
   Mean angular error 
    &17.21  &16.44  &16.60  &16.78  &17.66  &17.08  &16.05  &15.25
                            &16.89 \\
   Standard deviation of angular error
    & 7.52  & 7.44  & 7.52  & 7.52  & 7.38  & 7.31  & 7.80  & 7.78
                            &7.50  \\

  \end{tabular}
\end{table*}



To complement the data of Table~\ref{tab:results-real-data2}, we have
also evaluated the execution timing details of the evaluated algorithms,
with the results shown in Table~\ref{tab:results-time}.

Table~\ref{tab:results-real-data2} shows that the \RobustDenoisingAlg{}
algorithm performs better than the \emph{x-mult} algorithms. Note that
this happens even when the input data is the same, and neither the
geometry of the array, nor the distance to the source $r$ are used for
denonising in our proposal. In what respect to computational demands,
our proposal is over 50 times faster.

Comparing with the \emph{b\&b} algorithm, our proposal is close to its
performance, which is also an important result provided that (again) we
do not use the array geometry for denoising, and our execution time is
over 110 times faster.

\begin{table}[t!]
  \renewcommand{\arraystretch}{1.3}
  \caption{Average execution time (s) and standard deviation for each
    method (evaluated on 9435 trials).}
  \label{tab:results-time}
  \centering
  \begin{tabular}{c c c c}
    \hline
                  &\RobustDenoisingAlg{} &\emph{x-mult}  &\emph{b\&b} \\
    \hline
   Mean time (s) &0.024&1.255&2.649\\
   Std (s)       &0.0011&0.086&0.339\\
    
  \end{tabular}
\end{table}
\section{Conclusions}
\label{sec:conclusions}

This paper has studied the properties of TDOA matrices, showing that
they can be effectively used for solving TDOA denoising problems. In
particular, the paper has investigated challenging scenarios where the
TDOA matrix is contaminated with Gaussian noise, outliers and where a
percentage of the measurements are missing. The paper shows that
denoising in the presence of Gaussian noise and missing data can be
solved in closed-form. This result is important, as it is the basis of
an iterative algorithm that can also cope with outliers. The paper has
tested the proposed algorithms in the context of acoustic localization
using microphone arrays. The experimental results, both on real and
synthetic data have shown that our algorithms successfully perform
denoising (up to 30\% of improvement in localization accuracy) with a
high rate of missing data (up to 50\%) and outliers, without knowing
the sensor positions. This is important as it opens its
application to tasks where the sensors geometry is
unknown. Interestingly, in real datasets our robust denoising
algorithm is systematically better than the Gauss-Markov estimator
even when there is no missing data.  This is also an important result as it
proves that the assumption of Gaussian noise does not hold in real
cases, while our robust model is capable of automatically discard
erroneous measurements. The proposed robust denoising method has also
been compared with other methods in literature on a localization
task. Our results are very similar to the state of the art, even
though we do not require knowing the array geometry in the denoising
stage.  Furthermore, our proposal is significantly less
computationally demanding. 

As for future work, we plan to further test
our denoising algorithms in applications where the position of the
sensors is unknown in advance, such in self-localization and
beamforming.

\section*{Acknowledgements}
\label{sec:acknowledgements}

This work has been supported by the Spanish Ministry of Economy and
Competitiveness under project SPACES-UAH (TIN2013-47630-C2-1-R), and by
the University of Alcalá under projects DETECTOR and ARMIS. Afsaneh
Asaei is supported by funding from SNSF project PHASER, grant agreement
number 200021-153507. We specially thank the anonymous reviewers for
their careful reading of our manuscript and their many insightful
comments and suggestions.

\bibliographystyle{IEEEtran}
\bibliography{Sparse_SRP,TDOA_Denoising,calibration}
%

\end{document}